\documentclass{aa}  

\usepackage{graphicx}
\usepackage{txfonts}
\usepackage{calligra}
\usepackage[T1]{fontenc}
\usepackage[mathcal]{eucal}
\newcommand{\Msun}{${\rm M}_{\odot}$\ }
\newcommand{\Ft}{${F}(t)$\ }

\begin{document}

   \title{Observational signatures of microlensing in gravitational waves at LIGO/Virgo frequencies}


   \author{J.M. Diego
         \inst{1}\fnmsep\thanks{jdiego@ifca.unican.es}
          \and
         O.A. Hannuksela\inst{2}
          \and
         P.L. Kelly\inst{3}
          \and
         T. Broadhurst\inst{4,5,6}
          \and
         K. Kim\inst{2}
          \and
         T.G.F. Li\inst{2}
          \and
         G.F. Smoot\inst{7}
          }

   \institute{Instituto de F\'isica de Cantabria (CSIC-UC). Avda. Los Castros s/n. 39005 Santander, Spain
         \and
         Department of Physics, Chinese University of Hong Kong, Shatin, New Territories, Hong Kong
         \and
         School of Physics and Astronomy, University of Minnesota, 116 Church Street SE, Minneapolis, MN 55455, USA
         \and
         Department of Theoretical Physics, University of The Basque Country UPV/EHU, E-48080 Bilbao, Spain
         \and
         IKERBASQUE, Basque Foundation for Science, E-48013 Bilbao, Spain
         \and
        Donostia International Physics Center (DIPC), 20018 Donostia, The Basque Country
         \and
         Institute for Advanced Study, The Hong Kong University of Science and Technology, Clear Water Bay, Kowloon, Hong Kong
             }


 \abstract{
   Microlenses with typical stellar masses (a few ${\rm M}_{\odot}$) have traditionally been disregarded as potential sources of gravitational lensing effects at LIGO/Virgo frequencies, since the time delays are often much smaller than the inverse of the frequencies probed by LIGO/Virgo, resulting in negligible interference effects at LIGO/Virgo frequencies. While this is true for isolated microlenses in this mass regime, we show how, under certain circumstances and for realistic scenarios, a population of microlenses (for instance stars and remnants from a galaxy halo or from the intracluster medium) embedded in a  macromodel potential (galaxy or cluster) can conspire together to produce time delays of order one millisecond which would produce significant interference distortions in the observed strains. At sufficiently large magnification factors (of several hundred), microlensing effects should be common in gravitationally lensed gravitational waves. We explore the regime where the predicted signal falls in the frequency range probed by LIGO/Virgo. We find that stellar mass microlenses, permeating the lens plane, and near critical curves, can introduce interference distortions in strongly lensed gravitational waves. For those lensed events with negative parity, (or saddle points, never studied before in the context of gravitational waves), and that take place near caustics of macromodels, they are more likely to produce measurable interference effects at LIGO/Virgo frequencies. This is the first study that explores the effect of a realistic population of microlenses, plus a macromodel, on strongly lensed gravitational waves.   
  }
   \keywords{gravitational lensing --
                gravitational waves -- microlensing -- dark matter -- 
                cosmology
               }

   \maketitle
%

\section{Introduction}
Gravitational-wave astronomy has recently become a reality with the first detection of gravitational waves (GW hereafter) by the LIGO and Virgo ground-based interferemeters. To date, 11 events have been reported by the LIGO and Virgo detectors \citep{LIGO2018} and this number will quickly increase to tens of events in the coming years. Some of these events may correspond to gravitationally lensed events with magnification factors ranging from a few tens to a few hundreds \citep{dai2017effect, Ng2018, li2018gravitational, Smith:2017mqu, smith2018strong, Broadhurst2019}. Recent works have studied lensing effects in the existing LIGO/Virgo O1 and O2 events~\citep{hannuksela2019search, Broadhurst2019}, while~\citep{smith2018deep} searched for candidate galaxy cluster lenses for the the GW170814 event. The most likely lenses for such events would be massive galaxies or galaxy clusters~\citep{Ng2018, dai2017effect, smith2018strong, Broadhurst2019}. On the other extreme of the lens mass regime, compact objects with masses of a few hundreds to a few tens \Msun can also act as lenses~\citep{Lai2018}. In this case, the geometric optics limit is not valid since the Schwarszchild radius of the lens is comparable to the wavelength of the wave. For these relatively low masses, the lensing effect has a modest impact on the {\it average} magnification, but it can introduce a frequency dependence on the magnification \citep[see for instance][]{Jung2019,Lai2018}.  
An even smaller mass regime was considered in \cite{Christian2018} where the authors find that lenses with a mass as low as 30 \Msun could be detected with current experiments. They consider also future, higher-sensitivity experiments and show how they can push the limit to even smaller masses of order 1 \Msun.  These conclusions are, however, obtained assuming isolated microlenses and without accounting for the effect of the macromodel, or other nearby microlenses. 

The signal-to-noise (SNR) of and observed GW that is being lensed with magnification $\mu$ scales (to first order) as SNR$\sim \sqrt{\mu}/D_l(z)$, where $D_l(z)$ is the luminosity distance to the GW source at redshift $z$. If we denote by $D_m$ as the maximum luminosity distance at which a GW with a given chirp mass can be observed (that is, with SNR equal to the minimum SNR for detections, SNR$_{min}$), then the SNR of a wave that originates at this maximum distance (or equivalenty at the maximum redshift $z_m$) and that is magnified by a factor $\mu$ would be equal to SNR$_{min}\sqrt{\mu}D_m/D_l(z)$. It then follows that an event with magnification $\mu$ can be observed up to a maximum luminosity distance $D_l(z)=\sqrt{\mu}D_m$. Even for a modest value of $\mu\approx4$, the volume from which GWs can be observed would increase by almost an order of magnitude. 

For low redshifts of the GW, the probability of lensing is low. This probability is often referred to as the optical depth of lensing, $\tau$, and to first order it can be interpreted as the integrated (over the mass function of lenses) cross section of all gravitational lenses from redshift 0 to the redshift of the source. In the simplest calculations, the cross section of an individual lens is given simply by its Einstein radius. A more detailed calculation of the optical depth incorporates the profiles of the lenses as well as their ellipticity and computes the probability of lensing in the source plane by inverse ray shooting techniques \cite[see for instance][]{Diego2018b}. 

Independently of the models assumed for the lenses, the optical depth is found to increase quickly between source redshift 0 and source redshift 1. Beyond redshift 1, the optical depth keeps increasing but a slower pace, and beyond redshift $\approx 3$ the increase in optical depth is only at the percent level compared with the optical depth for a source at $z=3$. Hence, one would expect gravitational lensing of GW to be more much common for sources beyond $z=1$ than for sources below $z=1$. As an example, the optical depth for magnification factors $\mu>4$ and for sources at redshift $z=2$ is found to be around $10^{-4}$ \citep[see for instance][]{Turner1984,Fukugita1992,Kochanek1996,Hilbert2008,Takahashi2011,Diego2018b}, meaning that approximately 1 in 10000 GWs at $z=2$ would be magnified by a factor $\mu>4$. If one considers a spherical shell of thickness  $\delta z=0.2$ around $z=2$, one finds a volume of $\approx 100\  {\rm Gpc}^3$. If one considers the local rate of BBH inferred by LIGO ($\approx 200\, {\rm Gpc}^{-3} {\rm yr}^{-1}$) and extrapolates this rate to $z=2$ assuming an evolution model that follows the star formation rate \citep[that is, a rate a factor $\approx 4$ times larger at $z=2$ than at $z=0$, ][]{Madau2014}, one finds that, in the shell volume considered above, there is approximately $10^5$ events per year, out of which around 10 events would be amplified by a factor $\mu>4$ (some of these events will be multiplied multiple times increasing the number of detections). \cite{Ng2018} finds estimates for the number of lensed events, arriving at a total of $\sim 1-100$ lensed events per year at design sensitivity. An experiment that has the sensitivity to detect unlensed events up to $z=1$, could potentially detect these lensed events at $z=2$. Future studies will be able to estimate the number of these types of lensed events more accurately. 

The next observing runs will see further sensitivity upgrades to both LIGO and Virgo, as well as the prospects of a fourth detector called KAGRA~\citep{somiya:2012detector,aso2013interferometer,akutsu2018construction} joining the network. Simultaneously, another detector is being built in India~\citep{M1100296}. As the sensitivities of these ground-based detectors improve in the near future, GW experiments will reach the z=1 frontier, opening the door to study lensed GW events.
These lensed events will be interesting for multiple reasons. The most obvious one is that they will allow to extend the range of redshift at which GWs can be detected and set strong constraints on the evolution of events as a function of redshift (the number of observed lensed events is proportional to the unknown rate at high redshift times the {\it known} optical depth of lensing). As interesting as this might be, this paper focuses on a different aspect. GWs that are lensed by factors larger than a few, will inevitably travel through areas near the critical curves of the lenses. The size of these critical curves normally trace the distribution of matter of the lenses in regions where the convergence is $\approx 0.5-0.8$. At these convergence values, the presence of numerous microlenses along the path of the GW is not only possible but unavoidable \citep[see for instance][]{Diego2018a}. These microlenses can be stars and remnants from the galaxy or intracluster medium but also more exotic microlenses such as the hypothesized primordial black holes (PBH). Although numerous studies show that PBH can not account for all dark matter, in the mass regime of a few tens of solar masses, PBH can still account for a few percent of the total mass budget. Even a small fraction of dark matter in the form of PBH could account for the rate of binary black hole merger observed by LIGO/Virgo \citep[see for instance][]{Carr2017,Liu2018,Liu2019}. 
Whether or not LIGO/Virgo is observing a population of PBH with masses around 30 \Msun is still an open question. Microlensing can severely constrain the fraction of PBH in the universe \citep[see for instance][]{Diego2018a,Oguri2018}. Microlenses can introduce a modulation in the magnification of the GW as a function of frequency \citep{Deguchi1986,nakamura1998gravitational,takahashi2003gravitational}. 
Earlier work has suggested that microlenses with masses below 100 \Msun can not produce observable effects in GWs observed by LIGO/Virgo \citep[see however][where the authors claim LIGO/Virgo could detect masses as small as 30 \Msun if the signal to noise ratio is at least a factor 30]{Christian2018}. One of the reasons for this is due to the very small time delay between images produced by microlenses in this mass range. Such time delays are much smaller than the inverse of the frequency of the GWs observed by LIGO/Virgo, so no observable interference is produced at frequencies below 500 Hz. We argue that this is not necessarily the case and that in certain configurations, GWs can be significantly affected by microlenses down to $\approx 1$ \Msun.

Although microlenses have been considered in the context of GW, they are often considered as isolated point sources. This is certainly not the case when dealing with microlenses at cosmological distances and near critical curves\footnote{The approximation would be valid, however, for microlenses in our own Galaxy}. Instead, microlenses are always embedded in the potential of the macrolens (galaxy or cluster) they belong to. The effect of the macromodel can not be ignored as it can have a significant impact on the observed lensing signal. For instance, a point lens with no external shear will reduce its caustic to a single point and will produce only two counterimages. On the contrary, a microlens with an external shear will have a diamond shape caustic and typically produces 3 images. In the context of GWs, having two or three images can make a significant difference since the interference pattern can change substantially (two time delays instead of one). Moreover, when approaching the critical curve of a lens (that is, at larger magnification factors $\mu$), a fixed unit of area in the image plane gets compressed in the source plane by a factor $\mu$. This has important implications for the lensing of GWs. A large compression factor (i.e a large $\mu$) can result in overlapping microcaustics in the source plane \citep[see for instance][]{Kayser1986,Paczynski1986,Wambsganss1990,Wyithe2001,Kochanek2004,Diego2018a,Diego2018b}. These overlapping microcaustics can produce images that are separated by relatively large distances in the image plane, or in other words, can produce images with relative time delays which are larger than the time delays they would produce without overlap. If the time delay between two lensed waves is of the order of the inverse of the frequency of the GW, both waves interfere and the observed signal could show this interference. In this paper we explore the regime of GWs magnified by a macromodel in the presence of microlenses with stellar masses. 

The structure of this paper is as follows. In section \ref{sect_theoryI} we present the basic lensing formalism that is relevant for this paper (geometric optics limit). 
Section \ref{sect_theoryII} discusses the magnification when the geometric optics limit is not valid and one needs to consider wave optics. In section \ref{sect_Numerical} we discuss the algorithm used to compute the magnification in the wave optics regime and tests the algorithm performance with a simple example for which an exact analytical solution exists in section \ref{sect_ResultsI}. 
In sections \ref{sect_ResultsII}, \ref{sect_ResultsIII} and \ref{sect_ResultsIV} we present examples of magnifications as a function of GW frequency for a range of examples. Each section increases the degree of complexity of the microlens model. Section \ref{sect_ResultsII} starts with the simple case of a single microlens embedded in a macromodel potential. Section \ref{sect_ResultsIII} considers the particular case of two microlenses (with overlapping caustics), and with masses comparable to the massive black holes of the primaries that are being found by LIGO/Virgo. Section \ref{sect_ResultsIV} considers the realistic case of a population of microlenses with stellar masses but that are near a critical curve region, that is, it addresses the question of what is the role of microlenses in GW lensed events at large magnification factors. Finally, we discuss our results and conclude in section \ref{sect_concl}. In all cases, and without loss of generality, we assume a standard cosmological model with $\Omega_m=0.3$, $\Lambda=0.7$ and $h=0.7$ consistent with the latest constraints on the cosmological model from {\it Planck}. We also assume that the GWs originate at z=2 and that the microlenses (and macrolens) are at $z=0.5$. Our results depend weakly on these assumptions provided the lens is at cosmological distances and the GWs are at least twice that distance. 

\section{Deflection field for microlenses in macromodel potentials}\label{sect_theoryI}
In this section we briefly review the lensing formalism for microlenses embedded in a macromodel potential (a host galaxy or galaxy cluster). This topic has been widely covered in the literature \citep{Chang1979,Chang1984,Kayser1986,Paczynski1986}. 
For simplicity, we define the macromodel with just two parameters, the macromodel magnification factors in the radial and tangential direction, or $\mu_r$ and $\mu_t$ respectively. This simple choice for the macromodel is sufficient for our purpose since we are dealing with very small regions of the sky where the changes in the macromodel properties are insignificant. 
Without loss of generality, we assume that $\mu_t \gg \mu_r$ and that the main direction of the shear, $\gamma$, is oriented in the horizontal direction, that is $\gamma_2=0$ and $\gamma=\sqrt{\gamma_1^2 + \gamma_2^2}=\gamma_1$. 
Given $\mu_t$ and $\mu_r$, the corresponding values of $\kappa$ (convergence) and $\gamma$ (shear) can be found easily from the relation between $\kappa$, $\gamma$, $\mu_r$ and $\mu_t$. 
For a given choice of $\kappa$, and $\gamma$, the lens equation ($\vec{\beta} = \vec{\theta} - \vec{\alpha}(\vec{\theta})$) of the macromodel can be expressed as 
 \begin{equation}
 \vec{\beta}=  \vec{\theta} - \vec{\alpha}(\vec{\theta}) =
\begin{pmatrix} 1-\kappa-\gamma_1 & -\gamma_2 \\ -\gamma_2 & 1-\kappa+\gamma_1 \end{pmatrix} \vec{\theta},
\end{equation}
where the positions in the source plane are given by the coordinates $\beta=(\beta_x,\beta_y)$ and the positions in the image plane are given by the coordinates $\theta=(\theta_x,\theta_y)$.

The lensing potential of the macromodel, $\psi$, is given by 
\begin{equation}
\psi=\frac{\kappa}{2}(\theta_x^2 + \theta_y^2) + \frac{\gamma_1}{2}(\theta_x^2 - \theta_y^2) - \gamma_2\theta_x\theta_y = \frac{\kappa}{2}(\theta_x^2 + \theta_y^2) + \frac{\gamma}{2}(\theta_x^2 - \theta_y^2)
\end{equation}
where $\theta_x$ and $\theta_y$ are given in radians and we ignore a constant additive term (i.e, the potential is identically zero at the origin of coordinates of $\theta$). 

Since both the deflection field and lensing potential are linear with the addition of new masses, if a population of $N$ point masses are present, the deflection, $\vec{\alpha}_{PS}(\vec{\theta})$,  and potential, $\psi_{PS}(\vec{\theta})$, from the distribution of point masses  can be simply added to the above equations with;  
\begin{equation}
\vec{\alpha}_{PS}(\vec{\theta})=\sum_i^N\frac{4GM_iD_i(z_l,z_s)}{c^2}\frac{\delta\vec{\theta}_i}{|\delta\vec{\theta}_i|^2},
\end{equation}
and,
\begin{equation}
\psi_{PS}(\vec{\theta})=\sum_i^N\frac{4GM_iD_i(z_l,z_s)}{c^2}ln(|\delta\vec{\theta_i}|),
\end{equation}
where $\delta\vec{\theta}_i=\vec{\theta}-\vec{\theta}_i$ is the distance to the point mass $i$ at $\vec{\theta}_i$ and with mass $M_i$, $D_i(z_l,z_s)$ is the geometric factor $D_i(z_l,z_s)=D_{ls}(z_l,z_s)/(D_l(z_l)D_s(z_s))$ with $D_{ls}(z_l,z_s)$, $(D_l(z_l)$ and $D_s(z_s))$ the angular diameter distances between the lens and the source, between the observer and the lens, and between the observer and the source respectively. 

A quantity of interest, that will be relevant in section \ref{sect_crowded}, is the effective optical depth, $\tau_{eff}$ introduced by \cite{Diego2018a},
\begin{equation}
\tau_{eff}=(4.2\times 10^{-4})\Sigma\frac{\mu}{\mu_r}
\end{equation}
where $\Sigma$ (expressed in units of $M_{\odot}/pc^2$ in the expression above) is the microlens surface mass density. When $\tau_{eff}\approx 1$, the saturation regime is reached. In this regime, caustics constantly overlap in the source plane, and any source moving across a field with $\tau_{eff} > 1$ will always be experiencing microlensing \citep{Diego2018a,Diego2018b}. Since typical values for $\Sigma(M_{\odot}/pc^2)$ range between a few to a few tens, and assuming a typical value for $\mu_r \sim 1$, it is clear from the expression above, that the saturation regime is reached when the macromodel magnification is in the range of a few hundred to a few thousand. Similar values for the macromdoel magnification have been observed already at the position of the microlensing events of the Icarus and Warhol high redshift stars \citep{Kelly2018,Chen2019}. Given the fact that GW experiments have access to the entire sky at any given moment (as opposed to the aforementioned examples of Icarus and Warhol where a significant luck-factor had to be involved) events with similar, or even more extreme, macromodel magnifications are expected \citep[see][for a detailed estimation of the probability of these events]{Diego2018b}. In those scenarios, we expect microlensing to play a significant role. 

Finally, the time delay is given by
\begin{equation}
\Delta T = \frac{1+z_l}{cD(z_l,z_s)}\left[ \frac{1}{2}|\vec{\theta}-\vec{\beta}|^2 - \psi \right] = \Delta T_{geom} + \Delta T_{grav}
\end{equation}
Where we have assumed that all point masses are at the same redshift in the lens plane so the factors  $D_i(z_l,z_s)$ are the same for all of them. 
The term $\Delta T_{geom}$ is known as the geometric time delay and the term $\Delta T_{grav}$ is known as the gravitational time delay. 

It is sometimes useful to express the time delay in dimensionless units. Usually this is done by re-scaling both the angular positions and potential by the Einstein radius, $\theta_E^2=(4GM/c^2)D(z_l,z_s)$. This redefinition of the time delay expression makes most sense when one is dealing with a single microlens since in this case the Einstein radius can be defined without ambiguity, but in general one can still set the Einstein radius to any arbitrary mass, or scale, and still redefine the time delay equation. 
\begin{equation}
\Delta T = \frac{4GM(1+z_l)}{c^3}\left[ \frac{1}{2}|\vec{x}-\vec{y}|^2 - \tilde{\psi} \right] = 
\frac{2R_s(z_l)}{c}\left[ 1\vec{x}-\vec{y}|^2 - \tilde{\psi} \right],
\end{equation}
where $\vec{x}=\vec{\theta}/\theta_E$, $\vec{y}=\vec{\beta}/\theta_E$, $\tilde{\psi}=\psi/\theta_E^2$, and $R_s(z_l)$ is the redshifted Schwarzschild radius of the lens. 
The first term sets the scale of the time delay for a given lens mass,
$4GM/c^3 = 1.97\, 10^{-5} (M/{\rm M}_{\odot})$ seconds. This simple scaling, shows that for GW with frequencies $\nu\sim 100$ Hz, interference and diffraction effects are expected for isolated microlenses with masses $> 500/(1+z_l)$ \Msun. Below this mass, the Schwarzschild radius of the lens is smaller than the wavelength of the GW (with $\nu\sim 100$ Hz) and diffraction effects are not important resulting in the GW not {\it seeing} the microlens. However, as discussed in the sections below, this minimum mass can be lowered substantially under certain conditions that can increase the time delay between multiple images. 
   
As shown in the following section, the quantity of interest for lensing of GW is the term $2\pi\nu\Delta T$. If one defines the dimensionless frequency $w=\nu 8\pi GM(1+z_l)/c^3=4\pi\nu R_s(z_l)/\lambda$, where $\lambda$ is the wavelength of the GW, then,  $2\pi\nu\Delta T = w\left[ 0.5|\vec{x}-\vec{y}|^2 - \tilde{\psi} \right]$.

\section{Lensing of gravitational waves at large macromodel magnifications}\label{sect_theoryII}
Lensing of GW has been studied in detail in the past \citep{Wang1996,Nakamura1998,Sereno2010} and more recently in \citep{Dai2017,Jung2019,Broadhurst2018,Lai2018,Dai2018,Christian2018,Oguri2018b,Smith2018}. In a broader sense, extensive work has been done in the context of wave optics, which is the appropriate regime for studying the lensing of GWs when the Schwarschild radius of the lens is comparable to the wavelength of the GW \citep{Nakamura1998,Nakamura1999,takahashi2003gravitational}. Also, in the context of astrophysics, femtolensing is formally identical to lensing of GWs but at much higher frequency, and hence, involving much smaller microlenses \citep{UlmerGoodman1995}. Finally, the wave optics regime is covered also in traditional textbooks like for instance \cite{SchneiderBook1992}. This section presents only a brief description of the formalism that is relevant for this paper. The reader is directed to the above literature for more details.

When a GW is amplified by a factor $\mu$, the observed SNR increases by a factor $\sqrt{\mu}$. If the Schwarzschild radius of the lens is much larger than the wavelength of the GW, geometric optics can be applied and the observed strain gets amplified by the same factor $\sqrt{\mu}$ independently of the varying GW frequency. However, when the wavelength of the GW is comparable to the Schwarzschild radius of the lens, interference between multiply lensed GWs can take place, since in this case the time delay between GWs is comparable to the inverse of the frequency of the GW. In this regime, one needs to consider wave optics and the magnification factor may be significantly different several cycles before the merging event and right before the merger event~\citep[e.g.][]{takahashi2003gravitational}. This would introduce a modulation in the observed strain as a function of time that can be measured and used to infer the masses of the intervening lens~\citep{Cao2014,Lai2018}. Interference produced by multiply lensed images has been studied in detail in the literature. In the context of LIGO/Virgo observations, the mass of the lenses considered in earlier work is typically above 100 \Msun \citep[see however][]{Christian2018}. At smaller masses, isolated microlenses can produce time delays that are very small (much smaller than the inverse of the frequency of the GW) so interference does not take place. Also, much of the earlier work has studied the case of isolated microlenses (i.e with circular critical curves and point-like caustics), or at most microlenses with an external shear of moderate strength \citep{Lai2018}. To the best of our knowledge, no work has considered the case of multiple microlenses nor the regime where the potential from the macromodel can produce large magnification factors. Both scenarios can produce interesting effects that can be observed at LIGO frequencies. Large magnification factors are expected for observed GWs originating at redshift $z>1$. As discussed in \cite{Broadhurst2018}, observed lensed GW events are expected at large magnification factors (tens to few hundreds) for some models. A distribution of microlenses with stellar masses (mass $\sim 1$ \Msun) can produce overlapping caustics in the source plane. An event taking place inside one of these overlapping regions would appear in the image plane at separations similar to the separation between the microlenses. For instance, if two overlapping microcaustics originate from two microlenses with stellar masses but separated by several microarcsec, they could mimic the separation between microimages produced by a much larger microlens. 
Also, even for isolated microlenses, a large macromodel magnification can introduce a large geometric time delay, so even moderate masses can produce measurable interference effects. More precisely, a microlens with mass $M$ embedded in a macromodel with magnification $\mu$ behaves like a microlens with an effective mass $M\mu$ \citep{Diego2018a,Diego2018b}. 

   \begin{figure} 
   \centering
   \includegraphics[width=9.0cm]{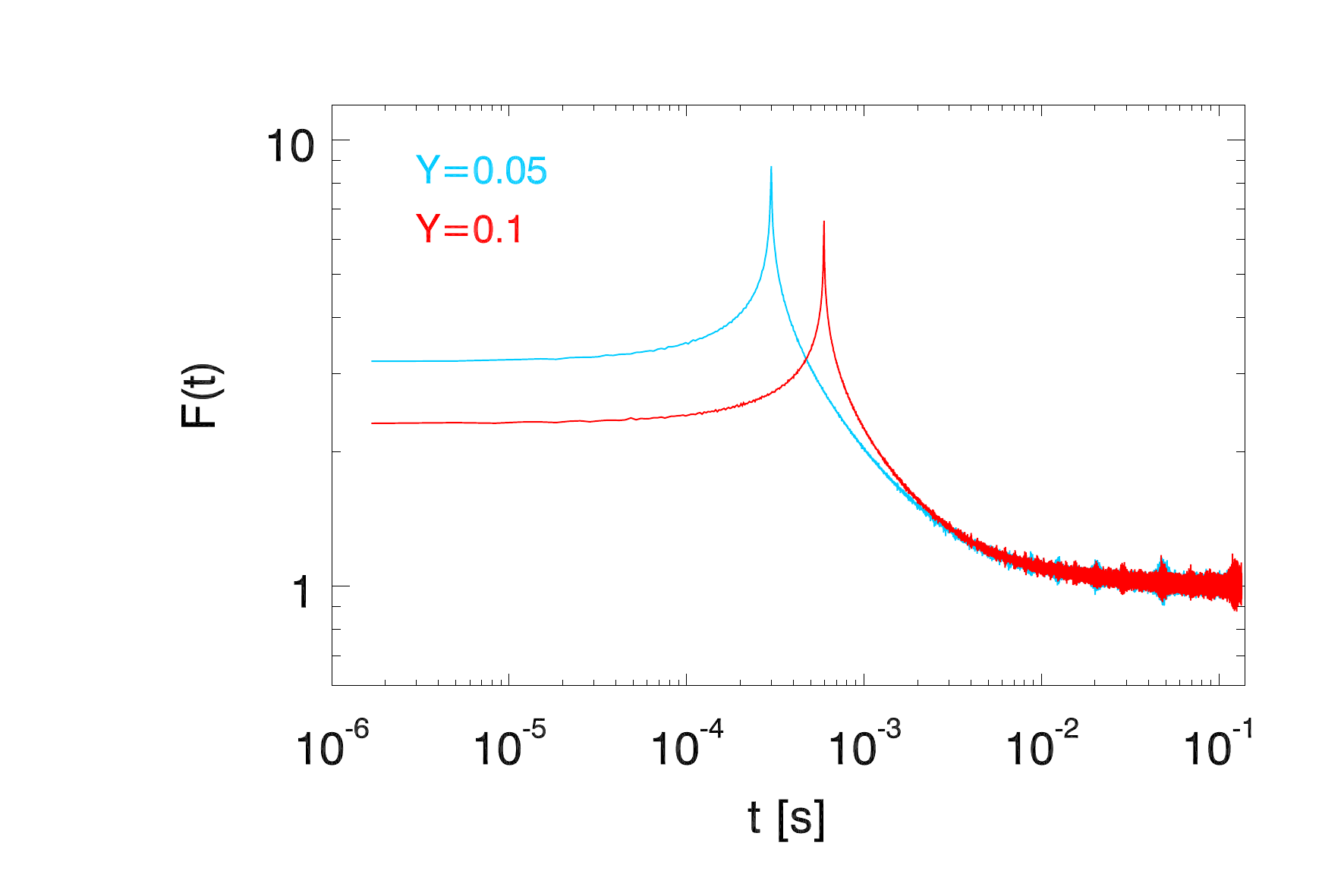}
      \caption{Function ${\rm F}(t)$ for an isolated point mass with $100$ \Msun and at a distance from the singular caustic $Y=\vec{\beta}/\Theta_E=0.05$ (blue solid line) and $Y=0.1$ (red solid line). Both curves are normalized to 1 at large time delays. 
              }
         \label{Fig_1PS_Ft}
   \end{figure}

   \begin{figure} 
   \centering
   \includegraphics[width=9.0cm]{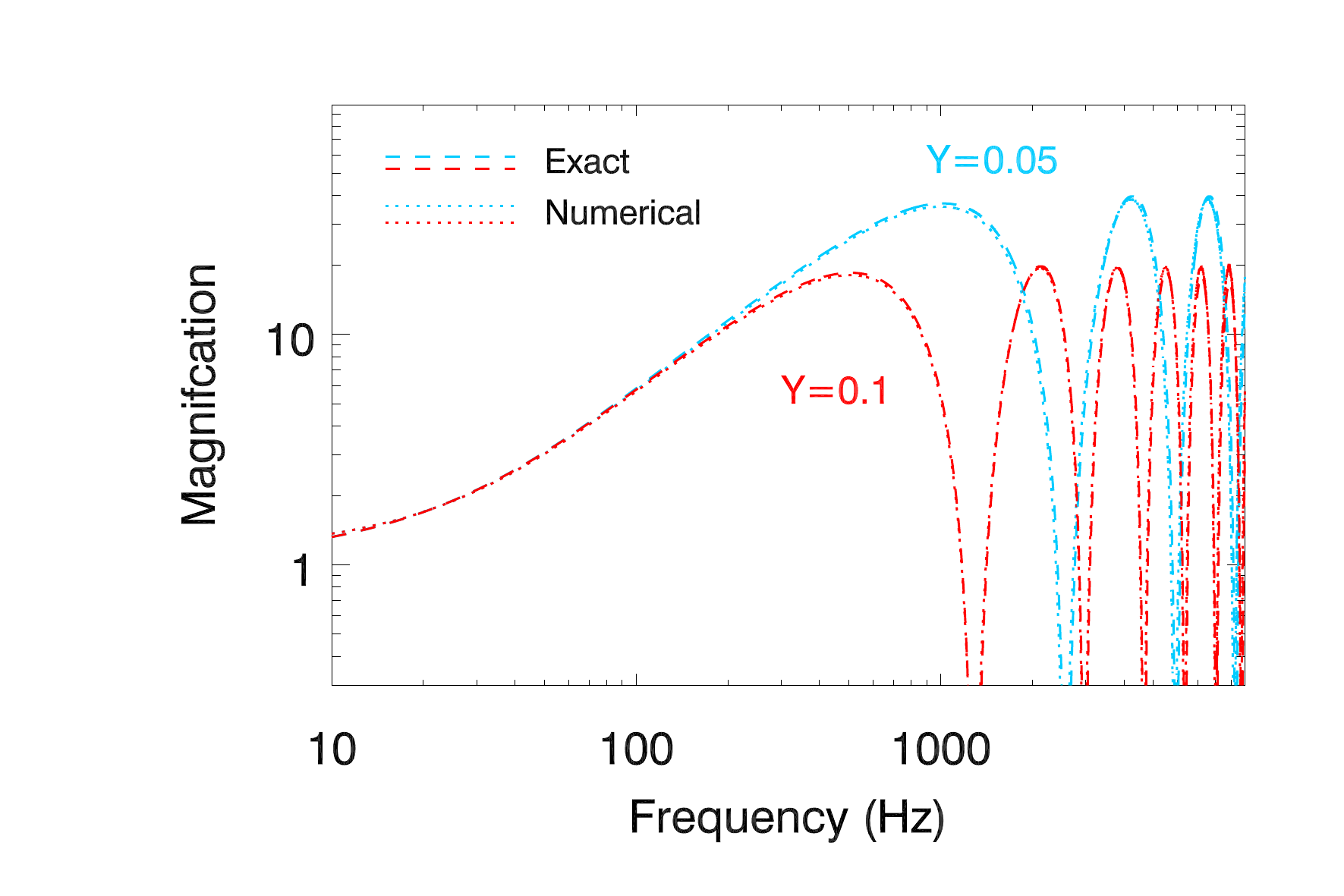}
      \caption{Predicted magnification for an isolated point mass with $100$ \Msun and at a distance from the singular caustic $Y=\vec{\beta}/\Theta_E=0.05$ 
               (blue lines) and $Y=0.1$ (red lines). 
               The dashed lines show the analytical (exact) prediction. The dotted lines show the result obtained with the numerical algorithm. 
              }
         \label{Fig_1PS_Fw}
   \end{figure}

The magnification in wave optics is given by the diffraction integral \citep[see for instance][]{SchneiderBook1992}. 

\begin{equation}
\sqrt{\mu} = |F(w,\beta)| = A_o \left | \frac{\nu}{2\pi i} \int d^2\theta\, e^{i2\pi\nu\Delta T(\theta,\beta)} \right |
\label{Eq_Fw}
\end{equation}
where $\nu$ is the frequency of the GW in Hz and the normalization $A_o$ guarantees that at very high frequencies one recovers the  geometric optics magnification when averaged over a frequency range. $\Delta T$ is the time delay between lensed images expressed in seconds. 

For simple models, like for instance an isolated point-like microlens, simple expressions for $\Delta T$ can be found and Eq.~\ref{Eq_Fw} can be evaluated analytically, yielding~\citep{Nakamura1998,takahashi2003gravitational}:
\begin{equation}
F(w,y) = e^{h(w,y)}\,\Gamma\left(1-i\frac{w}{2}\right)\, _1F_1\left( i\frac{w}{2},1;i\frac{wy^2}{2} \right )
\label{Eq_Fw_PS}
\end{equation}
where we remind the reader that $y=\beta/\theta_E$ and $w=(8\pi GM(1+z_l)/c^3)\nu$. The function $\Gamma$ is the standard gamma function and $_1F_1(a,b;z)$ is the Kummer's confluent hypergeometric function of the first kind. The term in the exponential part is given by:
\begin{equation}
h(w,y) = \frac{\pi w}{4}+i\frac{w}{2}\left[ {\rm ln}\left (\frac{w}{2} \right ) - \frac{\left ( \sqrt{Y^2+4}-Y \right )^2}{4} + {\rm ln}\left (\frac{\sqrt{Y^2+4}+Y}{2}\right ) \right]
\end{equation}
where $y=\beta \theta_E$. 

When multiple microlenses are present, the time delay between multiple images can adopt a complex form, but the integral can still be solved numerically as described in the next section.

   \begin{figure*}
   \centering
   \includegraphics[width=9.0cm]{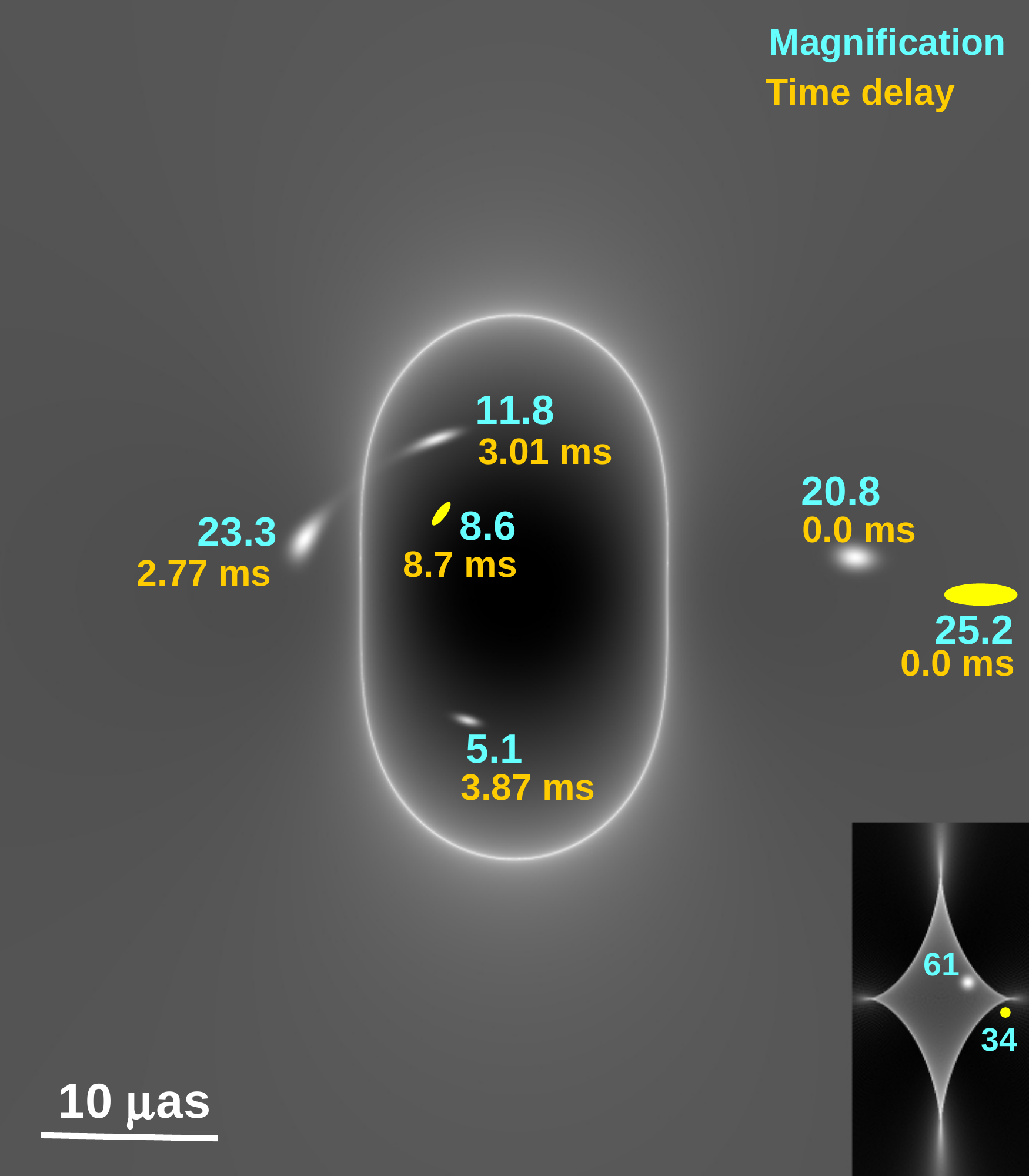}
    \includegraphics[width=9.0cm]{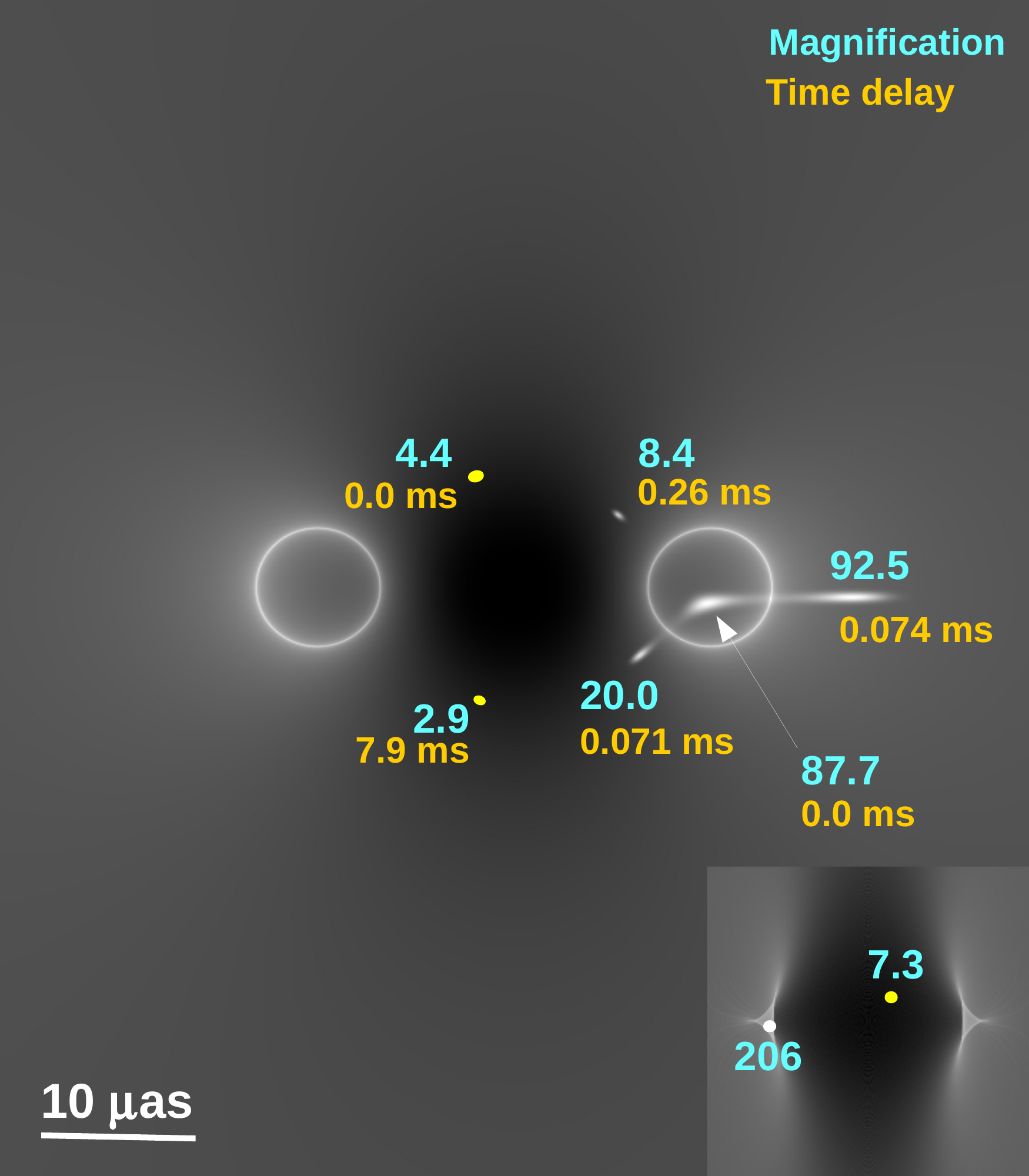}
     \caption{Example of a source at $z=2$ lensed by a microlens with 100 \Msun\,  embedded in a macromodel at $z=0.5$ with magnifications $\mu_t=10$ and $\mu_r=3$.
               {\it Left panel}. The large image shows the magnification around the microlens in the image plane (with the critical curve) and the microimages. The small image on the bottom-right shows the 
               magnification in the source plane with two sources in white and yellow. The surface brightness of the source is assumed to follow a Gaussian distribution (in white). 
               Without the microlens, there would be a single lensed microimage with magnification 30. The microlens boosts the magnification by a factor $\approx2$ and  
               breaks the microimage into four microimages introducing time delays between them of up to 3.87 ms, i.e in LIGO's range. 
               {\it Right panel}. Similar to the left panel but for a microlens in the side with macromodel negative parity. Note how the high magnification region is now smaller but the maximum magnification can be higher. In contrast, areas of small relative magnification (and larger time delays) are more common. 
              }
         \label{Fig_CC_Caustics_Case2}
   \end{figure*}

\section{Numerical evaluation of the observed magnifications}\label{sect_Numerical}
To evaluate Eq.~\ref{Eq_Fw} in a general situation, we follow \cite{UlmerGoodman1995}. The basic idea is to divide the integral in Eq.~\ref{Eq_Fw} in areas of equal time delay. Then it is easy to see that Eq.~\ref{Eq_Fw} is equivalent to computing the Fourier transform of $F(\Delta T)$, where $F(\Delta T)$ accounts for the area with a given time delay, $\Delta T$. For simplicity, we rename $F(\Delta T)=F(t)$. In \cite{UlmerGoodman1995}, the authors propose an algorithm where \Ft is computed as contour integrals. 
While that is a valid approach, we use the simpler, and faster, method of computing directly the area in constant intervals of $\Delta T$. 
\begin{equation}
F(t)=\int dxdy \delta(\Delta T(x,y)-t).      
\end{equation}

This method avoids also the divergence of the contour integrals at the local minima and maxima of the time delay function. Finally, the direct computation of the area from the time delay maps is equally time consuming for one, or many, microlenses, something that is not true for the countour integral method which first needs to identify contours around each microlens and later perform the contour integral along each contour.


\section{Performance of the numerical method. Isolated point masses}\label{sect_ResultsI}

To test the validity of the numerical approximation, we compute the lensing magnification as a function of the GW frequency for the particular case of an isolated microlens. For this case, an exact solution exists and it is given by Eq.~\ref{Eq_Fw_PS}. We simulate the deflection field around a microlens with mass 100 \Msun at redshift $z=0.5$. We place a source at redshift $z=2$ and at  distances 0.05 and 0.1 times the Einstein radius of the lens (that is, $Y=\beta/\theta=0.05$ and $Y=0.1$ in Eq.~\ref{Eq_Fw_PS}). For each configuration, we compute the time delay and derive the function $F(t)$.

These are shown in Figure~\ref{Fig_1PS_Ft}. The blue curve is for the case where $Y=0.05$ while the red curve is for $Y=0.1$. In both cases, the curves are normalized to 1 at large time delays. This is equivalent to dividing by the not-lensed signal, for which the corresponding $F(t)$ would be a constant. The $F(t)$ curves converge to the no-microlensing case at large time delays (that is, $F(t)=Cte$) and exhibit a logarithmic pulse at the corresponding time delay between the two counterimages \citep[(the logarithmic shape is discussed in more detail in][]{UlmerGoodman1995}. 

The amplitude of the logarithmic pulse is related with the magnification. As expected, the case with $Y=0.05$ has larger magnification but smaller time delay between the images than the case with $Y=0.1$. Fourier transforming back the $F(t)$ curves, and multiplying by the frequency one obtains the corresponding $F(w)$ \citep{UlmerGoodman1995,Nakamura1999}. In order to correct for the limited time span covered by $F(t)$, we apply an apodization (in real space) at large time delays to the curve $F(t)$. Without this apodization, the magnification at low frequencies ($\nu < 100$ Hz) tends to zero. We find that a standard cosine apodization function works remarkably well.   
The result is shown in Fig.~\ref{Fig_1PS_Fw}. The dashed lines show the results derived with the exact calculation. Again, the case with $Y=0.05$ attains larger magnifications than the case with $0.1$. In both cases, the average (over a wide range of frequencies) of the magnification converges to the magnification predicted by geometric optics. At low frequencies, both curves converge to 1 as expected since the corresponding wavelengths become smaller than the Schwarzschild radius of the lens. The dotted lines show the numerical approximation. The numerical approximation reproduces very well the exact solution with deviations at the sub-percent level.  At low frequencies, the GW does not {\it see} the microlens and the magnification tends to one. 

\section{Microlens embedded in a macromodel potential}\label{sect_ResultsII}
In the previous section we showed the result for an isolated lens and compared it with the analytical exact solution. That situation is unrealistic in the sense that most microlenses will be forming part of a halo (galaxy, group of galaxies or cluster). This halo, or macromodel, imprints a magnification in the GW which is generally insensitive to the frequency of the GW. That is, the effect of the macromodel over the GW can be treated in the geometric optics limit since the Schwarzschild radius  of the halo is orders of magnitude larger than the wavelength of the GW. For simplicity, we assume that the macromodel forms two counterimages. Although in most situations, a macromodel will form more than two images, usually two of them carry the bulk of the magnification and will be the most likely to be observed if the merger occurs at large redshift. However, our results can be applied to any of the counterimages. As mentioned above, these two macroimages can be treated in the geometric optics limit. We neglect here the very particular (and extremely rare) case where the final moments of the inspiral phase takes place very close (thousands of km) to one of the caustics of the macro-model. In this case, diffraction effects from the macromodel become also important and the geometric optics limit is no longer valid. This case is studied in the literature \citep[see for instance][and references therein]{SchneiderBook1992}. Usually, the two counterimages with the largest magnification have opposite parity. From now on, we refer to the counterimages produced by the macromodel as macroimages and we distinguish between the macroimage with positive parity and the macroimage with negative parity. The images produced by the microlens are referred to as microimages. If a macroimage intersects a microlens, the macroimage will, in general, break into multiple microimages. A macroimage with positive parity will break into microimages with both positive and negative parity. The situation is similar for macroimages with negative parity. 

\subsection{Magnification}

   \begin{figure}
   \centering
   \includegraphics[width=9.0cm]{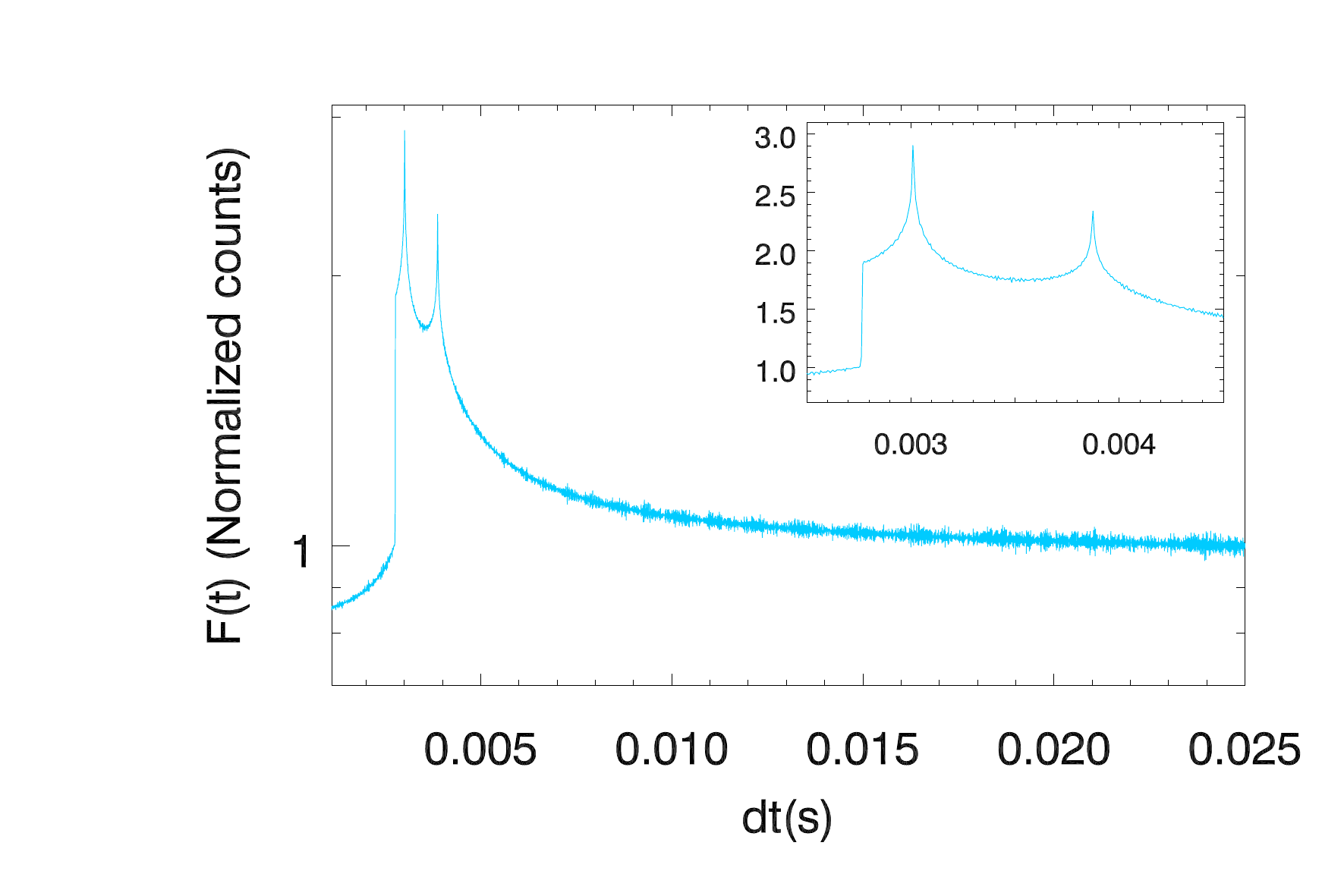}
      \caption{$F(t)$ for the white source shown in the left panel of Fig.\ref{Fig_CC_Caustics_Case2}. The zoom-in shows an enlarged version of the curve around the maxima. Note how the positions of the discontinuities and the logarithmic divergences correspond with the times shown in Fig.\ref{Fig_CC_Caustics_Case2}. 
              }
         \label{Fig_Ft_POS}
   \end{figure}

Before studying the interference pattern produced by the multiple microimages, it is illustrative to visualize the configuration of microimages in the geometric optics limit. To simulate microlenses in an external shear we follow \cite{Diego2018a}. 
For the macromodel, we adopt the values $\mu_t=10$ and $\mu_r=3$ which results in a net magnification of $\mu=30$. 
In general, the macromodel can impact substantially how the magnification behaves around a microlens. The only case where the effect of the macromodel on the microlenses can be safely ignored is for microlenses in the Milky Way but we do not treat this case here. Our focus is instead for microlenses in galaxies, groups or clusters at cosmological distances and GWs produced at redshifts of 1 or above. The caustic of a point-like microlens in the presence of an external shear (from the macromodel) is no longer a point but it adopts the familiar diamond shape typical of elliptical lenses if the microlens is on the side of the macromodel where images have positive parity (i.e, in a macrominima or macromaxima of the time delay). An example is shown in the left panel of Fig.~\ref{Fig_CC_Caustics_Case2}. The critical curves adopt the typical elongated shape of elliptical potentials. The small inset in the bottom-right corner shows the corresponding caustic in the source plane. The location of two sources are marked with a  white dot inside the caustic region and a yellow dot outside the caustic region. The number next to each source indicates the magnification in the source plane. The corresponding counterimages for each source are shown in the image plane (larger plot). For each microimage, we indicate the magnification in the image plane and the time delay (in milliseconds) with respect to the minimum of the time delay. The yellow microimages are shown with their approximate size and orientation. 
In the absence of microlens, the macromodel magnification would be $30$. When the microlens is introduced, the white source is magnified by almost a factor two times larger. The yellow source, on the other hand, has a magnification comparable to that of the macromodel. 
In the wave optics regime, the time delay between microimages plays an important role and can leave an observational imprints, even for groups of microimages that are unresolved. At time delays of order few millisecond, GW as those observed by LIGO/Virgo would interfere. As expected, the relative time delay between microimages near a critical curve is relatively short (0.24 milliseconds between the microimages with magnifications 23.2 and 11.8 respectively), but the time delay between the microimage at the minima (i.e at time delay 0) and the other microimages is in the regime where, at LIGO/Virgo frequencies, there could be observable effects. Even for the yellow source, one should expect a change in the magnification as a function of frequency at even lower frequencies, since the time delay is larger. In wave optics regime, the effect of microlenses extend farther away past the realm of the caustic region, increasing the probability of detecting a lensing event. 

On the side with negative parity (or saddle points of the time delay), the critical curves and the caustics split into two regions as shown in the right panel of Fig.~\ref{Fig_CC_Caustics_Case2}. The critical curves remain circular but the region of low magnification is now between the two critical curves. The region of low magnification is significantly larger than the region of high magnification which results in a higher probability of a GW being {\it demagnified} by the microlens than {\it magnified} by it \citep[see][for a more specific quantification of this difference]{Diego2018a,Diego2018b}. The caustics break into two separate triangular regions of higher magnification (relative to the macromodel magnification), with a large area in between them of lower magnification (relative to the macromodel magnification). As mentioned above, the area containing the lower magnification region is significantly larger than the area of higher magnifications. This fact has direct consequences on the probability of observing lensed events of compact sources (in the geometric optics limit) which are less likely on regions where the macromodel parity is negative. A good example is the Icarus event \citep[a highly magnified star at $z=1.49$, ][]{Kelly2018} which has been consistently observed on the side with positive parity but remains demagnified most of the time on the side with negative parity. In wave optics, as we noted earlier, at very low frequencies the waves are unaffected by the microlenses. Hence, traditionally the probability of detecting or not detecting an event has been considered to be independent of the macroimage parity. However, at sufficiently high frequencies, we show below how the waves interfere and create constructive and destructive patterns at frequencies in the LIGO/Virgo window. This results in a modulation of the strain as a function of the frequency. If this modulation is not accounted for in the templates used in the matched filtering searches for GW, they could go unnoticed. 

   \begin{figure}
   \centering
   \includegraphics[width=9.0cm]{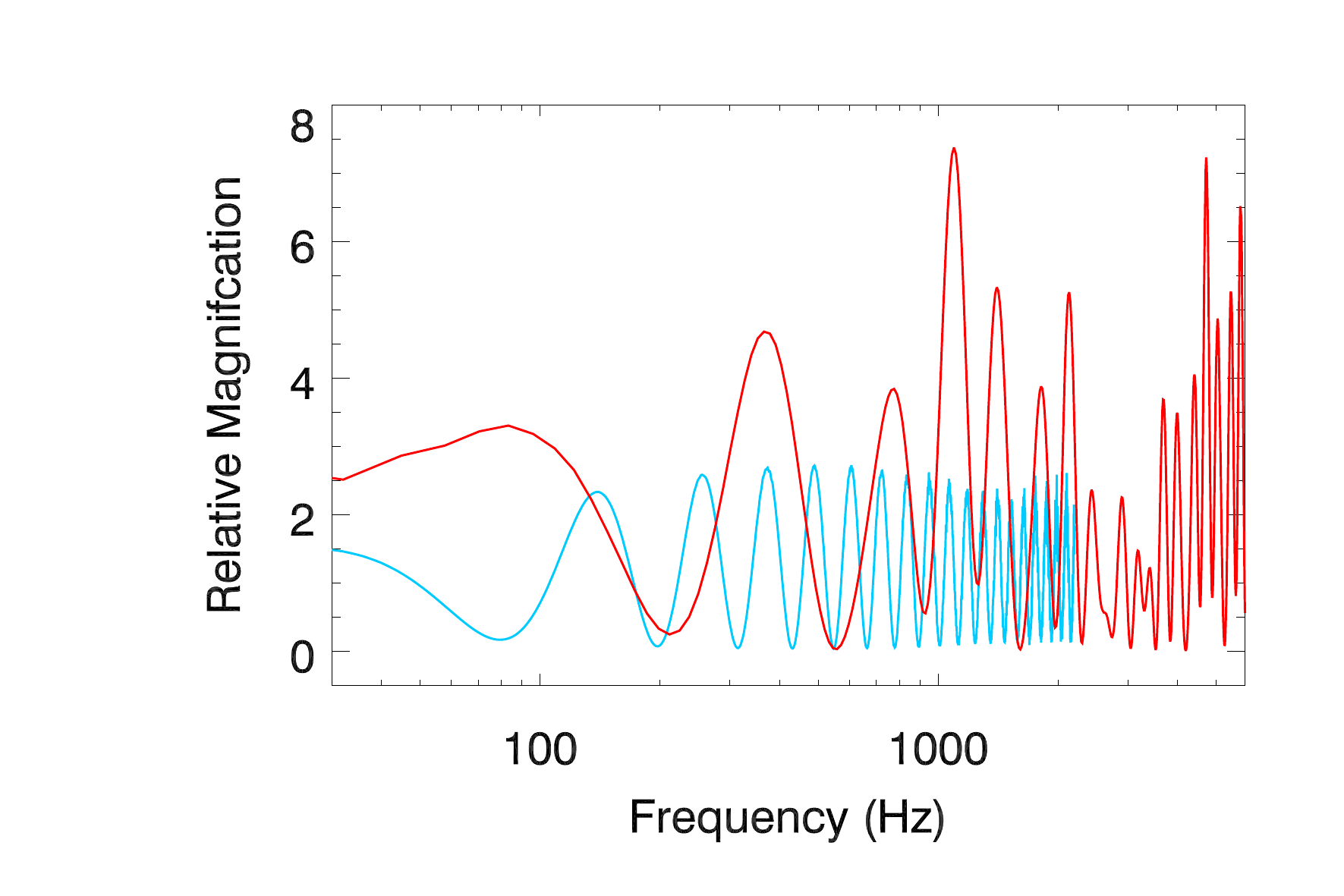}
      \caption{Observed magnification of the GW as a function of GW frequency for the two sources shown in the left panel of Fig.\ref{Fig_CC_Caustics_Case2}. The red curve is for the white source while the blue curve is for the yellow source. 
              }
         \label{Fig_Fw_POS}
   \end{figure}


The properties of caustics in regions of the macromodel with positive and negative parity have been studied in detail in the literature \citep[see for instance][]{Kayser1986,Paczynski1986,Wambsganss1990,Wyithe2001,Kochanek2004,Diego2018a,Diego2018b}. In the case of wave optics, this is not true (to the best of our knowledge) for microlensing of macroimages in regions with macromodel negative parity. The literature has focused so far only on microlensing of macroimages forming in macromodel regions with positive parity only, neglecting the impact that microlenses play on the macroimages forming in the regions with negative macromodel parity. In this work we study the two cases (parities) separately.  

\subsection{Microlensing of macroimages with positive parity}
This case has been studied in detail in previous work \citep{UlmerGoodman1995,Nakamura1998,Nakamura1999,Takahashi2003}. Here we consider the particular case of a microlens with 100 \Msun near a microimage that is being magnified  by a macromodel by a factor $30$. At this magnification, the signal to noise of the GW would be boosted by a factor $\sqrt{30}\approx 5.5$. With this magnification, an event at $z\approx 1.9$ with a chirp mass $M_c\approx 17$ \Msun would have been detected with the same significance as GW170729 (at $z=0.48$ and with almost twice the chirp mass). At magnifications $\mu> 30$ one expects $\approx 0.2$ events per year between $z=1.9$ and $z=2.1$ (or $\sim 1$ events per year in the redshift interval $2<z<3$) if one naively assumes a model for the rate evolution of GW events that traces the star formation rate history and it is normalized to 200 events per year and Gpc$^3$ at $z=0$ (consistent with current limits from LIGO) \citep[see for instance ][for an estimation of the probability of lensing at $z\approx 2$ and for $\mu>30$]{Diego2018b}.

To compute the magnification as a function of frequency, we consider two GWs at the positions indicated by the white and yellow dots in the left panel of Fig.~\ref{Fig_CC_Caustics_Case2}. In Fig.~\ref{Fig_Ft_POS} we show, in blue, the $F(t)$ extracted from the time delay for the white source (the $F(t)$ for the yellow source resembles that of the isolated microlens studied earlier with a single logarithmic peak at the time delay between the two yellow microimages). The inset on the top-right corner shows a zoom-in around the peak of $F(t)$. 
The curve $F(t)$ has three scales shown in the zoom-in inset plot, each one corresponding to a different time delay, 2.77 3.01 or 3.87 milliseconds. The relative height of each feature is dictated by the magnification of each microimage, as originally shown by \cite{UlmerGoodman1995}. The first feature (at 2.77 miliseconds) is a discontinuity corresponding to a microminima or micromaxima. The other two features are logarithmic divergences for each one of the two saddle points.  

Using the curves $F(t)$, one can easily compute the magnification as a function of frequency. The result is shown in Fig.~\ref{Fig_Fw_POS}. The red curve shows the magnification for the source inside the caustic region (white source in left panel of Fig.~\ref{Fig_CC_Caustics_Case2}). The blue curve is for the source outside the caustic region (yellow source in left panel of Fig.~\ref{Fig_CC_Caustics_Case2}). In both cases, the magnification is relative to the macromodel magnification, so magnification $\approx 1$ in this plot is to be understood as total observed magnification $\approx 30$. 
In the frequency range of LIGO ($\sim 50-500$ Hz), both curves exhibit features in the magnification that can be measured in future observations allowing for the unambiguous identification of lensed events (which would be otherwise difficult to recognize). 
At high frequencies, the red curve converges to the geometric optics limit prediction, that is to a mean value $\approx 2$ times the magnification of the macromodel. However, at LIGO frequencies, the magnification can be up to six times larger than the magnification from the macromodel, boosting the detectability of the GW in a given frequency range. This is a consequence of the constructive interference that takes place at certain frequencies. In the case of the blue curve, an undulatory behaviour is observed even at lower frequencies (consequence of the larger time delay of 8.7 milliseconds). Since one counterimage is magnified by a factor 25.2 and the other by a factor $\approx 3$ times smaller, a complete cancellation of the signal can not take place but the interference is large enough to be detected with the proper template. A classic search of GWs using standard templates may miss these events if this type of interference pattern is not accounted for. 
However, candidates to lensed events could be identified based on their unusually high chirp mass and/or low redshift \citep{Dai2017,Broadhurst2018, hannuksela2019search, Broadhurst2019}.
Also, multiple image analysis can be performed by following the methods proposed by \cite{Haris2018}. 

   \begin{figure}
   \centering
   \includegraphics[width=9.0cm]{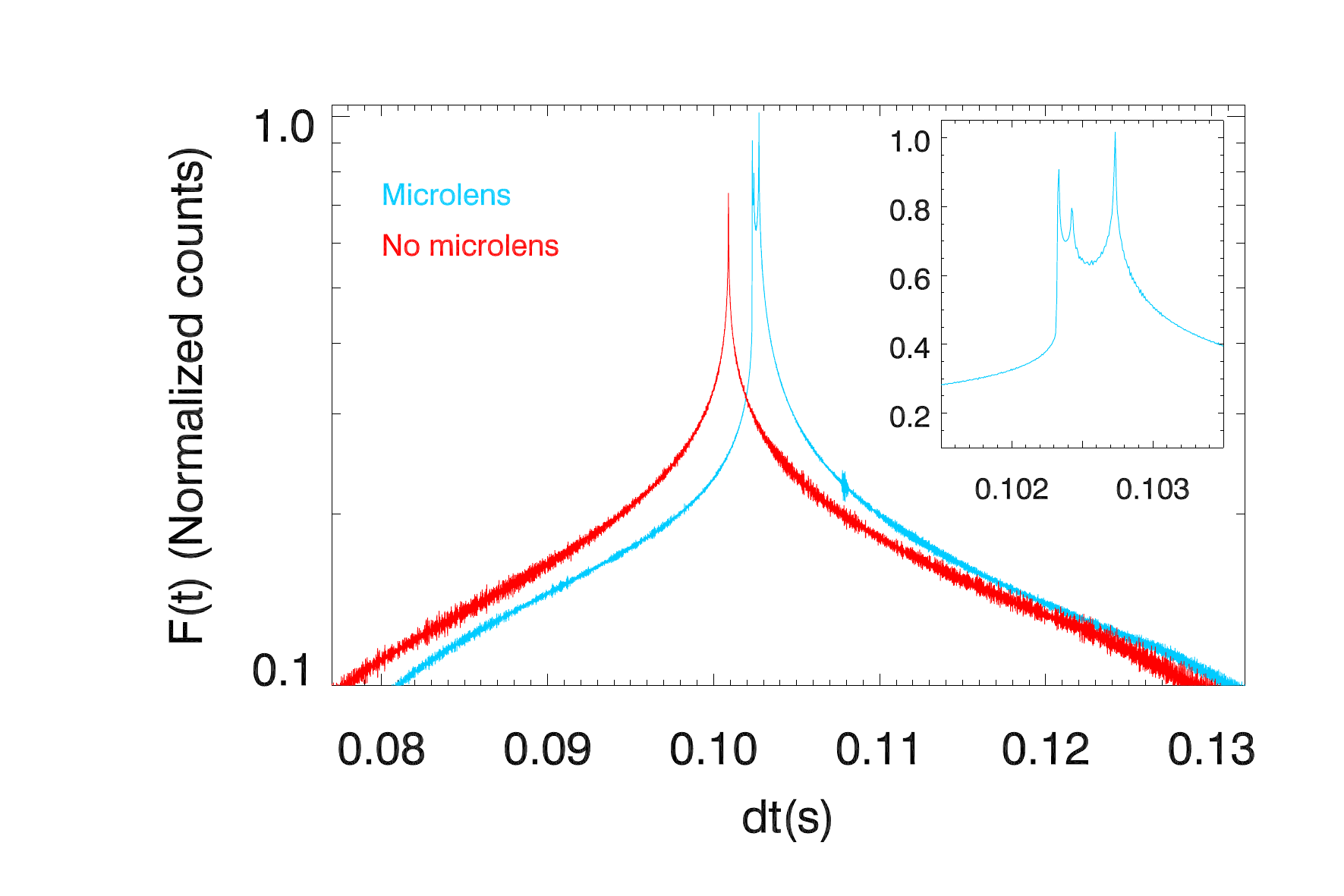}
      \caption{The blue curve shows the $F(t)$ curve for the white source shown in the right panel of Fig.~\ref{Fig_CC_Caustics_Case2}. The zoom-in shows an enlarged version of the curve around the maxima. The red curve is the corresponding $F(t)$ when the mass of the microlens is set to zero.
              }
         \label{Fig_Ft_NEG}
   \end{figure}

   \begin{figure}
   \centering
   \includegraphics[width=9.0cm]{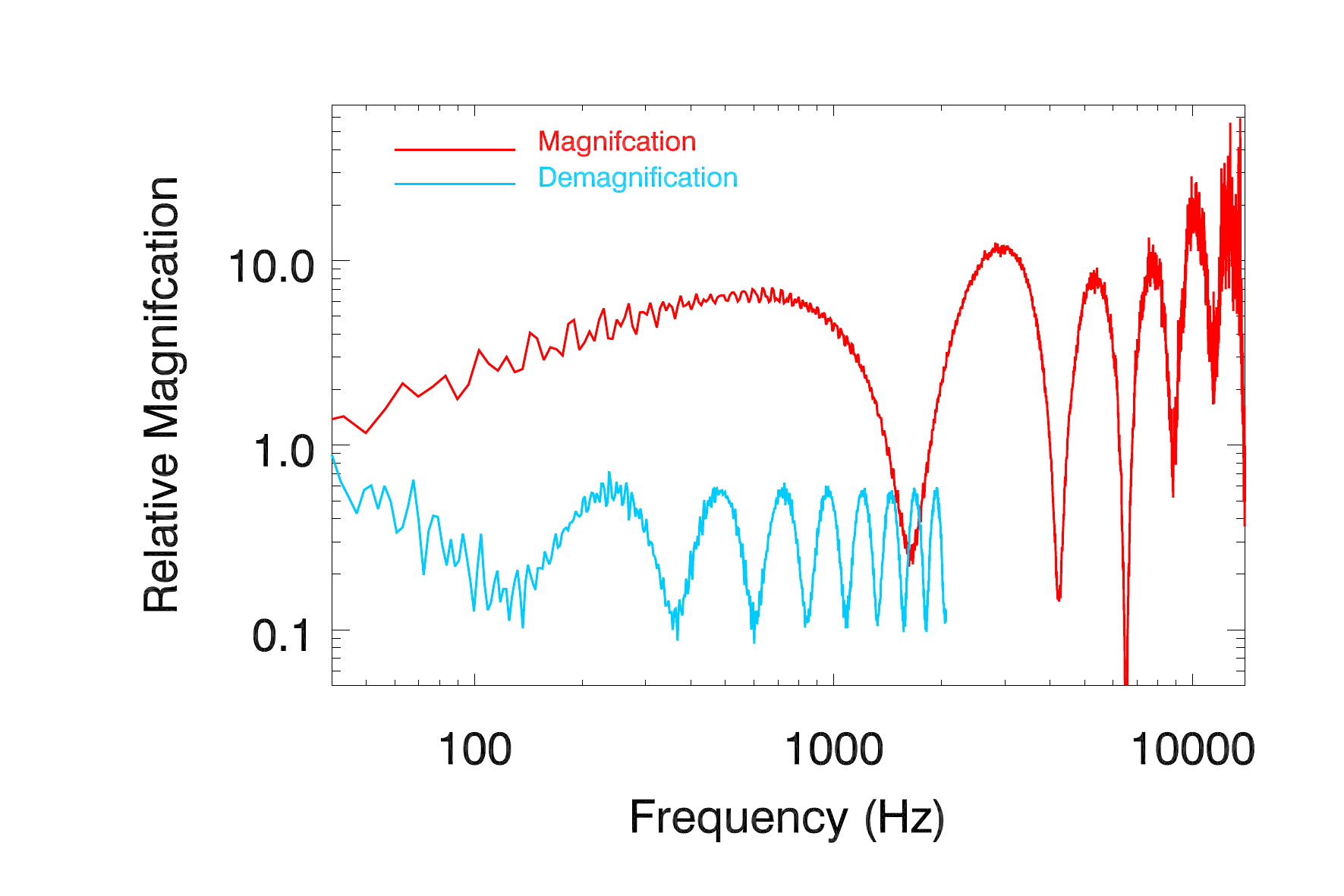}
      \caption{Observed magnification of the GW as a function of GW frequency for the two sources shown in the right panel of Fig.~\ref{Fig_CC_Caustics_Case2}. The red curve is for the white source (magnified with respect to the macromodel magnification) while the blue curve is for the yellow source (demagnified with respect to the macromodel magnification). Note how the blue curve converges to the smaller magnification value predicted in the geometric optics limit but it oscillates multiple times within the frequency range covered by LIGO.  
              }
         \label{Fig_Fw_NEG}
   \end{figure}


Finally, it is important to note that the frequencies shown in the x-axis of the previous figures, scale as the inverse of the mass of the microlens (this follows from the scaling of $\Delta T$ with the mass of the microlens). A mass 5 times smaller (i.e 20 \Msun) would still show significant fluctuations in the observed magnification within the frequency range covered by LIGO/Virgo. This result is in apparent disagreement with earlier work that concluded that only isolated microlenses above 100 \Msun could produce observable effects. However, that earlier work ignored the effect of the macromodel that helps boost the signal from the microlens. One should expect that at even larger magnifications, increasingly small masses may produce observable effects, similar to what happens in the geometric optics limit \citep[see][for a detailed discussion of this effect]{Diego2018a}. 

\subsection{Microlensing of macroimages with negative parity}
   \begin{figure*}
   \centering
   \includegraphics[width=9.0cm]{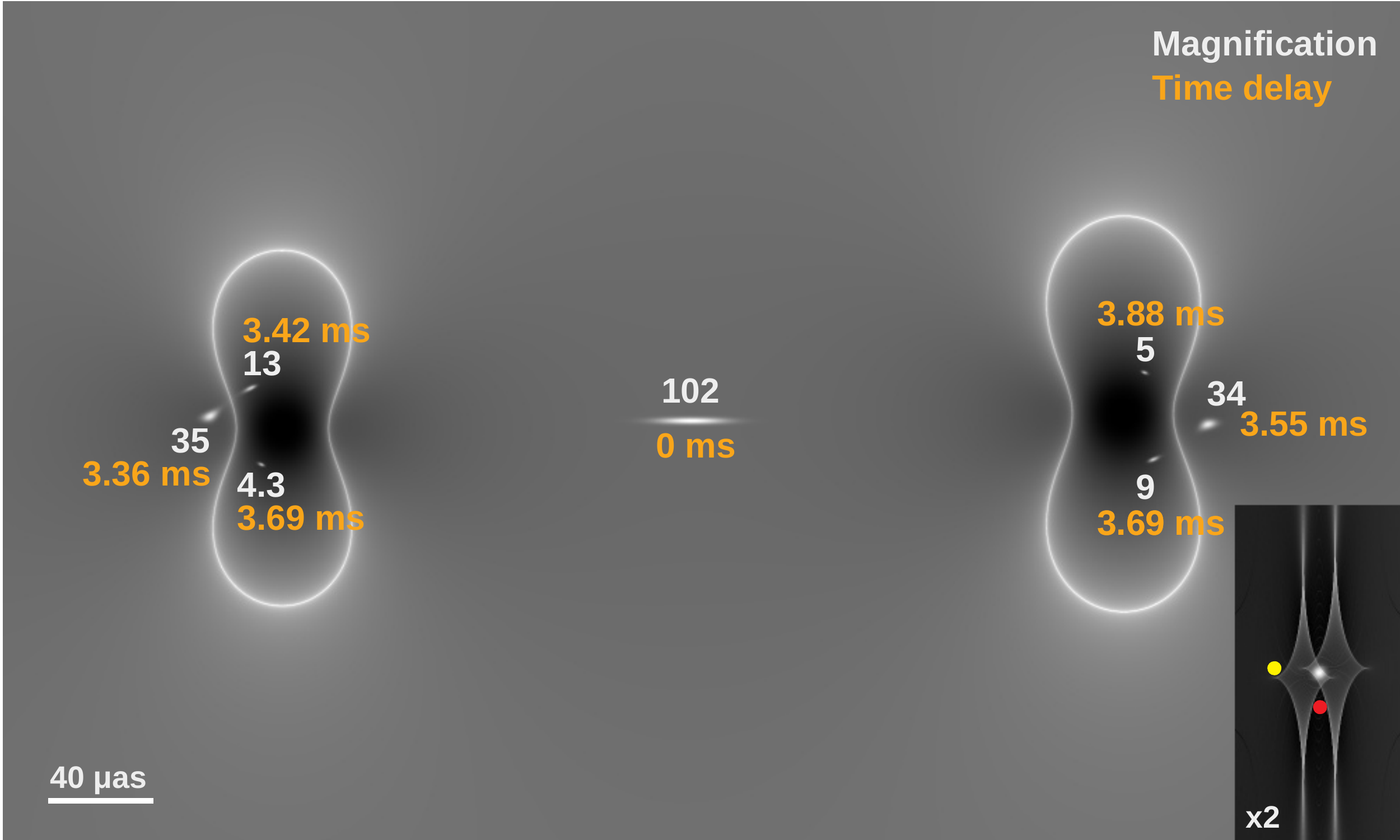}
   \includegraphics[width=9.0cm]{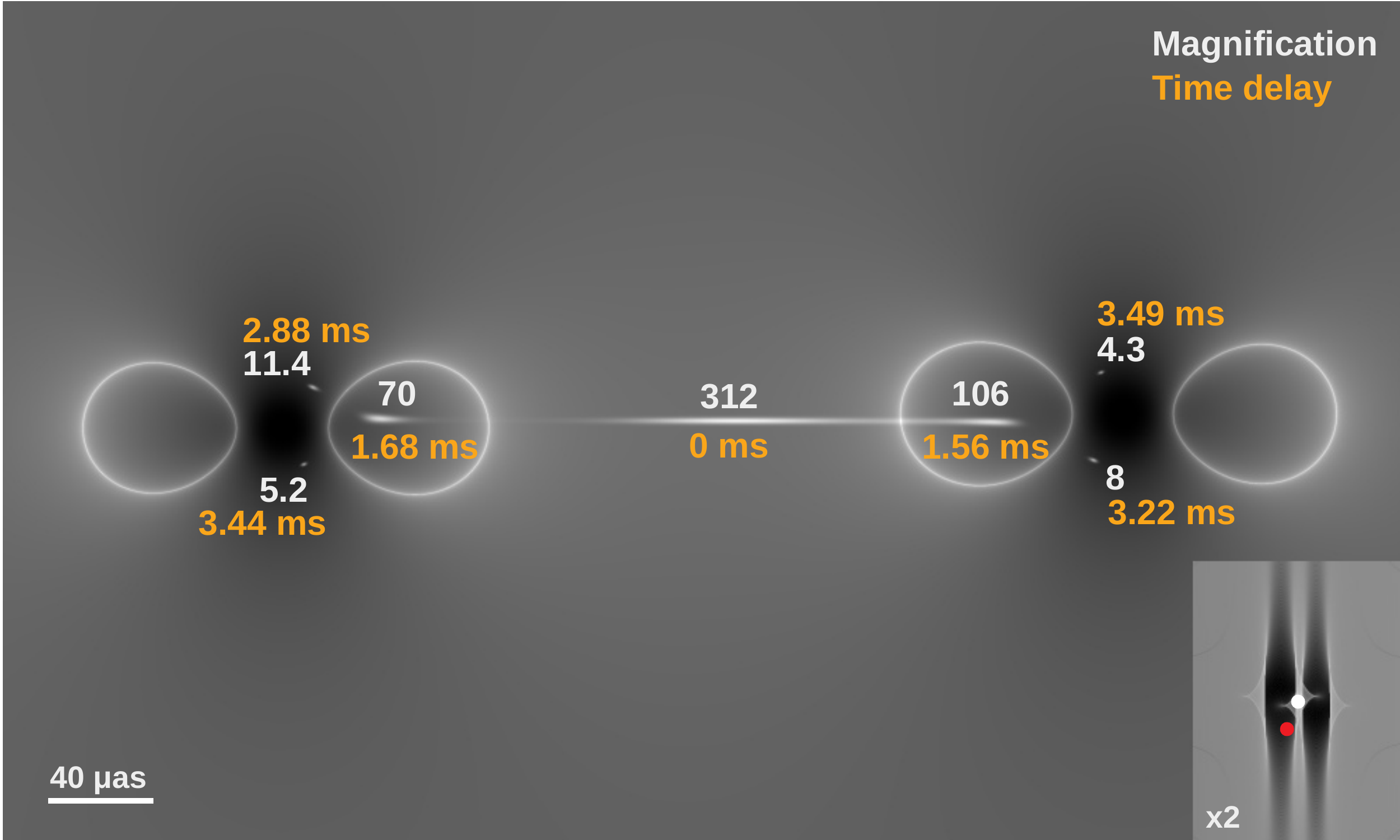}
      \caption{Like in  Fig.\ref{Fig_CC_Caustics_Case2} but for two microlenses with 25 \Msun and 30 \Msun. The left panel shows the case where the two microlenses are on a region of the macromodel with positive parity while the right panel is for when the same microlenses are in a region of the macromodel with similar magnification but negative parity. In both cases, the small figure in the bottom-right corner shows the source plane with the caustics and the location of the sources. Only the microimages produced by the white sources are shown in the bigger figures.   
              }
         \label{Fig_Arcs_TimeDelay_POS_NEG_3A}
   \end{figure*}
 In the absence of microlenses (or substructure, in general), macroimages with negative parity behave similarly to macroimages with positive parity in the geometric optics limit. However, when microlenses are present, the behaviour can change substantially  depending on the parity of the macroimage. When a macroimage intersects a microlens, the macroimage breaks into smaller microimages. The net magnification of the group of microimages may be comparable, larger or smaller than the macromodel magnification depending on where the GW is taking place, in relation to the microcaustics from the microlens. Contrary to the behaviour of macroimages, microimages can look very different depending on whether the microlens is on the side with positive or negative macromodel parity. In the particular case of microlenses on the side with macromodel negative parity, two microcaustics form with a gap of low magnification between them. The region of high magnification contained within the microcaustics can be much smaller than the area of low magnification in the gap. This results in a larger probability for microimages from a compact source to be demagnified if the corresponding macroimage is in a saddle point (or has negative parity) of the macromodel. We consider again a microlens with mass 100 \Msun embedded in a macromodel with magnification -30. This would correspond approximately to the conjugate position of the macroimage considered in the previous subsection.  The critical curves, and caustics are shown in the right panel of Fig.~\ref{Fig_CC_Caustics_Case2}. Note how the low magnification region between the two caustics is significantly larger than the large magnification region inside the two caustics. In the geometric optics limit, the larger probability of demagnification is compensated by a larger magnification inside the caustics region, so a moving source travelling across a microcaustic region would have similar integrated flux independently of whether the microlens is on the side with positive macromodel parity or the side with negative macromodel parity. The periods of maximum flux are shorter but more intense when the microlens is on the side with negative macromodel parity than when the same microlens is on the side with positive macromodel parity. On the other hand, the periods of low relative magnification (relative to the macromodel magnification) are more prolonged and at lower magnifications on the side with negative macromodel parity \citep[see more examples for varying magnifications in][]{Diego2018a,Diego2018b}. 

As in the previous subsection, we consider two sources that are shown in the inset small panel of the right side of Fig.~\ref{Fig_CC_Caustics_Case2}. 
The white source is inside the small caustic region and is highly magnified as shown in the source plane. The yellow source is in the more extended low magnification region and has a much smaller magnification. The configuration of the microimages, together with their corresponding magnifications and relative time delays, are also shown in Fig.~\ref{Fig_CC_Caustics_Case2}. For the white source, the relative time delays are significantly smaller than in the analogous case discussed in the previous section ($\approx 5$ times smaller). On the contrary, for the yellow source, the time delay is significant ($\approx 8$ milliseconds) and well within LIGO/Virgo frequency window. The curve $F(t)$ for the white source is shown as a blue curve in Fig.~\ref{Fig_Ft_NEG} (for the yellow source is again a simple logarithmic pulse). The blue curve shows one discontinuity corresponding to the microimage that arrives first and three logarithmic pulses for each one of the the other microimages. The red curve in the same plot is the $F(t)$ curve for the macromodel-only case, that is, without the microlens. In this case, the curve resembles the divergent logarithmic pulse typical of shallow points instead of a constant (like in the previous subsection).  

The magnifications are shown in Fig.~\ref{Fig_Fw_NEG}. The red curve is for the white source that is being magnified by the microlens and the blue curve is for the yellow source that is being demagnified by the microlens. The small scale fluctuations in these curves are artifacts in the numerical method. The red curve is almost featureless in the frequency range covered by LIGO/Virgo but it exhibits a continuously growing magnification up to $\sim 1000$ Hz that could be possible to be detected with future GW observations. On the contrary, the more likely event of the yellow source shows remarkable features that could be identified in future observations. At low frequencies, the magnification tends to the macromodel magnification but it falls sharply at around 100 Hz to then raise again and oscillate around a relative magnification of $\approx 0.25$ relative to the macromodel magnification. The fact that at low frequencies GWs are unaffected by microlenses could be used to identify GW in the sense that these waves could appear in matched filtering searches at lower frequencies but then disappear when including higher frequencies. Meanwhile, unmodeled searches may be able to identify these events even at higher frequencies~\cite[e.g.][]{2016PhRvD..93l2004A,2017PhRvD..95d2003A}. A dedicated search could be made after a highly magnified GW is detected since its corresponding counterimage (macroimage) should closely follow (or precede) the observed lensed GW. For highly magnified events, the time delay between macroimages should range between $\sim 1$ hour to days or months~\citep{Ng2018, Smith2018, Broadhurst2018} which should be useful to narrow the search for counterimages. As discussed in the earlier subsection, the time frequency scales with the inverse of the microlens mass so a microlens with mass as low as 20 \Msun would still be detectable through the interference signature.

\section{Pair of microlenses at large macromodel magnification}\label{sect_ResultsIII}
The examples studied above are of interest for situations where a massive microlens intersects the path of a GW. In this section we study an alternative scenario where two smaller microlenses overlap their caustics in the source plane. This situation interesting for large macromodel magnifications. As discussed earlier, a distant GW can be lensed by factors of a few tens with a non-negligible probability. Overlapping of caustics is not only possible, but unavoidable at large macromodel magnifications. The overlapping effect becomes increasingly more important as one approaches one of the critical curves of the macromodel. As the radial or tangential magnification approaches its diverging value (at the critical curve), a given number density of microlenses in the image plane (with their associated microcritical curves) results in a larger number density of microcaustics in the source plane. At some value of the macromodel magnification, regularly spaced microcritical curves in the image plane start to produce overlapping caustics in the source plane. A GW that intersects one of these overlapping regions will have its counterimages spread over the wider region covered by the corresponding microcritical curves, with the consequent increase in time delay. Small overlapping caustics can then mimic the effect of a much larger microlens (in terms of time delays), and hence produce effects at LIGO/Virgo frequencies. 

The number of possible combinations of microlens mass, distance between microlenses, and macromodel magnification is too large to be studied in detail. Here we present a simple example. We consider the special case where microlenses are primordial black holes (or PBH) with masses of a few tens of \Msun (although our example is valid also for regular black holes, like those being found by LIGO/Virgo). PBHs are a viable candidate to explain a fraction of the dark matter. Even though observations rule out the possibility of PBH explaining all the dark matter in a wide range of masses, there are some windows where fractions of the order of 10\% of the dark matter could still be in the form of compact objects, such as PBHs. One of these windows is for PBH masses of a few tens of \Msun \citep{Carr2017}. Interestingly, the latest results on gravitational waves from binary black hole (BBH) mergers could suggest a higher than expected abundance of black holes with masses of a few tens of \Msun \citep{LIGO2018}.

As above, we consider the two possibilities; macroimages forming on the side with positive parity and macroimages forming on the side with negative parity. 

\subsection{Macroimages with positive parity}
\begin{figure}
   \centering
   \includegraphics[width=9.0cm]{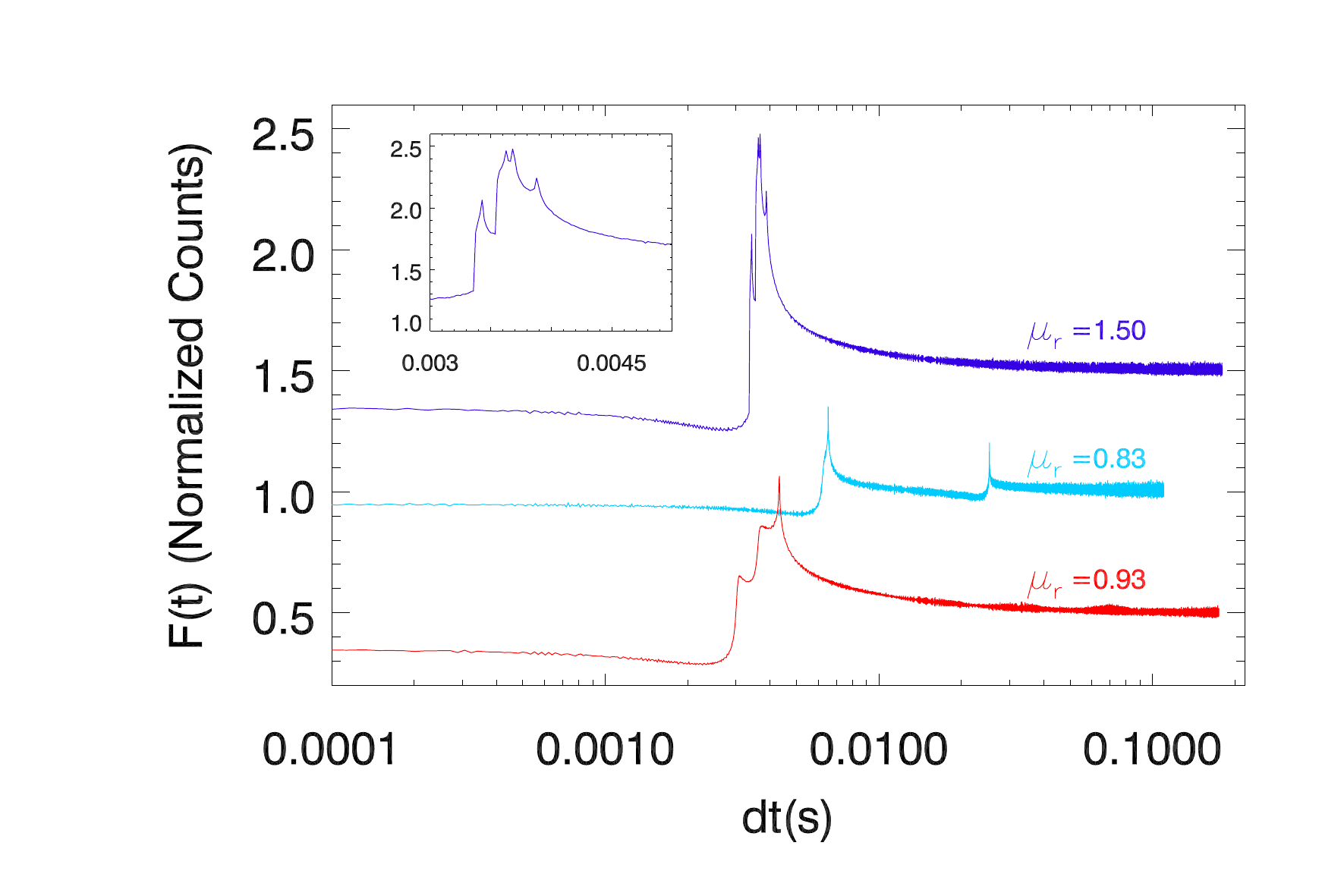}
      \caption{$F(t)$ curves for the three sources shown in the left panel of Fig.~\ref{Fig_Arcs_TimeDelay_POS_NEG_3A}. The dark blue curve is for the white source, the light-blue curve is for the yellow source and the red curve is for the red source. 
              }
         \label{Fig_Ft_POS_3A}
   \end{figure}

We place two point masses with 25 \Msun and 30 \Msun in a macromodel with magnification $\approx 150$ (or $50\times 3$). The two microlenses are separated by a distance of 320 $\mu as$. At this separation, the macromodel projects the critical curves into two overlapping caustics as shown in the small panel of the left Fig.~\ref{Fig_Arcs_TimeDelay_POS_NEG_3A}. The magnification of the macromodel is larger than in the previous sections by a factor 5. This translates into a probability 25 times smaller of having these events but the accessible volume increases by a factor $\sim 5$ (compared with the volume up to $z=2$) compensating partially the reduction in lensing probability. At large magnifications, the probability of microcaustics overlapping increases as discussed in detail in \cite{Diego2018a,Diego2018b}. 
An event in the  source plane that takes place in the caustic overlapping region will produce multiple images around both critical curves. Fig.~\ref{Fig_Arcs_TimeDelay_POS_NEG_3A} shows an example (in the geometric optics limit) of multiple images produced by a Gaussian source (depicted as a white round source in the small panel in the bottom right) that is placed in the overlapping caustic region. The total magnification in the geometric optics limit is $\approx 200$, that is $\approx 33\%$ times higher than the magnification that would be observed without the microlens. 
The microimage with the largest magnification is the one that arrives first and it appears in between the two critical curves. Smaller microimages form around each one of the critical curves. The time delays between the largest microimage and the other ones is of the order of a few milliseconds, sufficient to produce effects in the frequency range of LIGO/Virgo. Two of the smaller microimages are micro-minima or micro-maxima and the remaining four microimages are all saddle points. The $F(t)$ curve for this case is shown in Fig.~\ref{Fig_Ft_POS_3A} as a dark blue curve. The zoom-in shows the two discontinuities and the four logarithmic impulses for the two micro-minima or micro-maxima and the four saddle points. Two other curves are shown in this plot for two alternative locations of the source. These two sources are marked with a red and yellow dots in the bottom-right part of the left panel in Fig.~\ref{Fig_Arcs_TimeDelay_POS_NEG_3A}. The light-blue $F(t)$ curve corresponds to the yellow dot and the red  $F(t)$ curve corresponds to the red dot. Note how in the light-blue curve, a saddle point microimage forms at relatively large time delays ($dt\approx 25$ milliseconds). The microimages for these alternative locations are not shown in Fig.~\ref{Fig_Arcs_TimeDelay_POS_NEG_3A} for simplicity purposes. 

   \begin{figure}
   \centering
   \includegraphics[width=9.0cm]{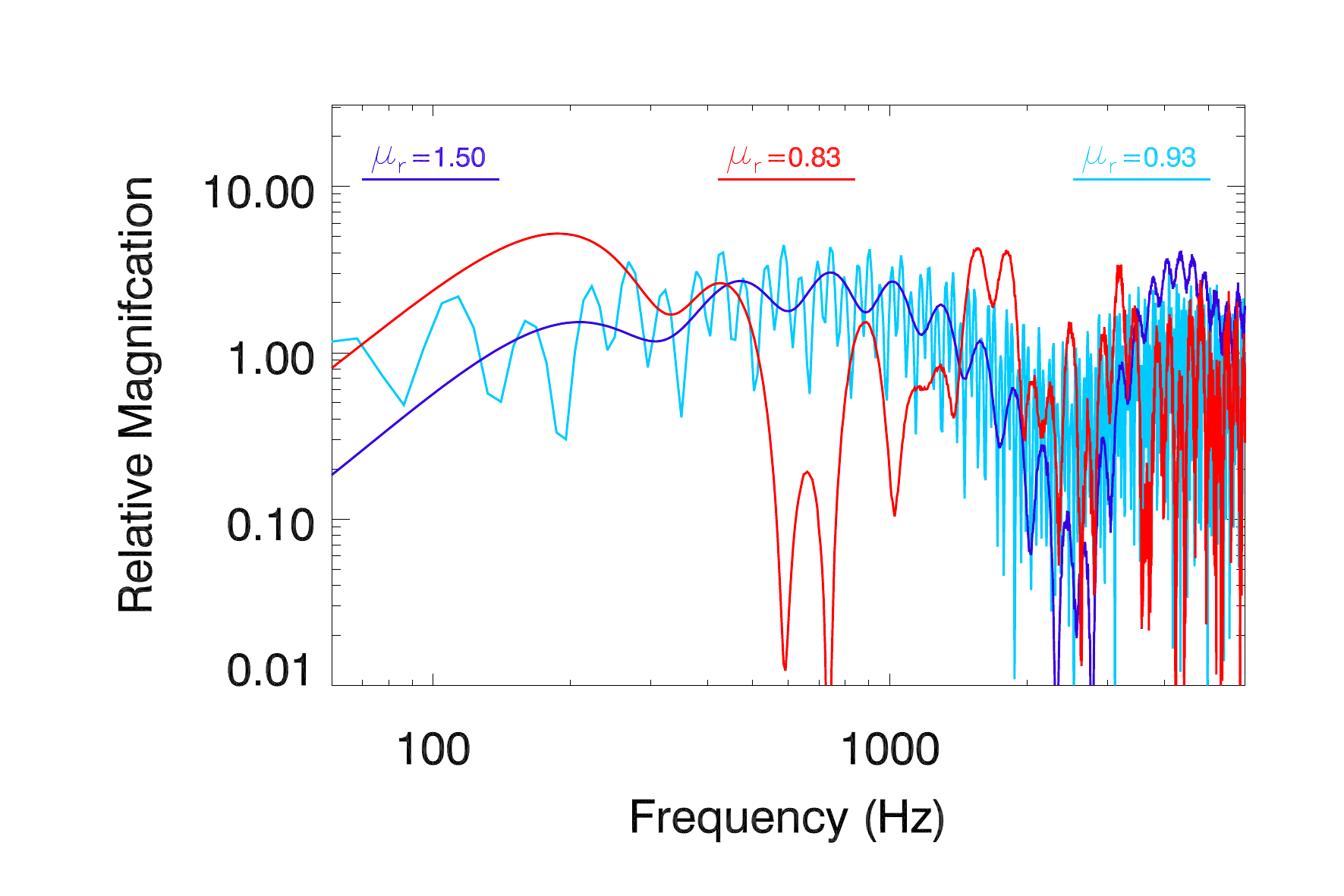}
      \caption{Magnification for the three sources shown in the left panel of Fig.~\ref{Fig_Arcs_TimeDelay_POS_NEG_3A} as a function of frequency. The color coding is the same as in Fig.~\ref{Fig_Ft_POS_3A}. The rapid oscillations in the light-blue curve are real and due to the peak at large time delays ($\approx 25$ milliseconds) shown in the light-blue curve of Fig.~\ref{Fig_Ft_POS_3A}.
              }
         \label{Fig_Fw_POS_3A}
   \end{figure}
   \begin{figure}
   \centering
   \includegraphics[width=9.0cm]{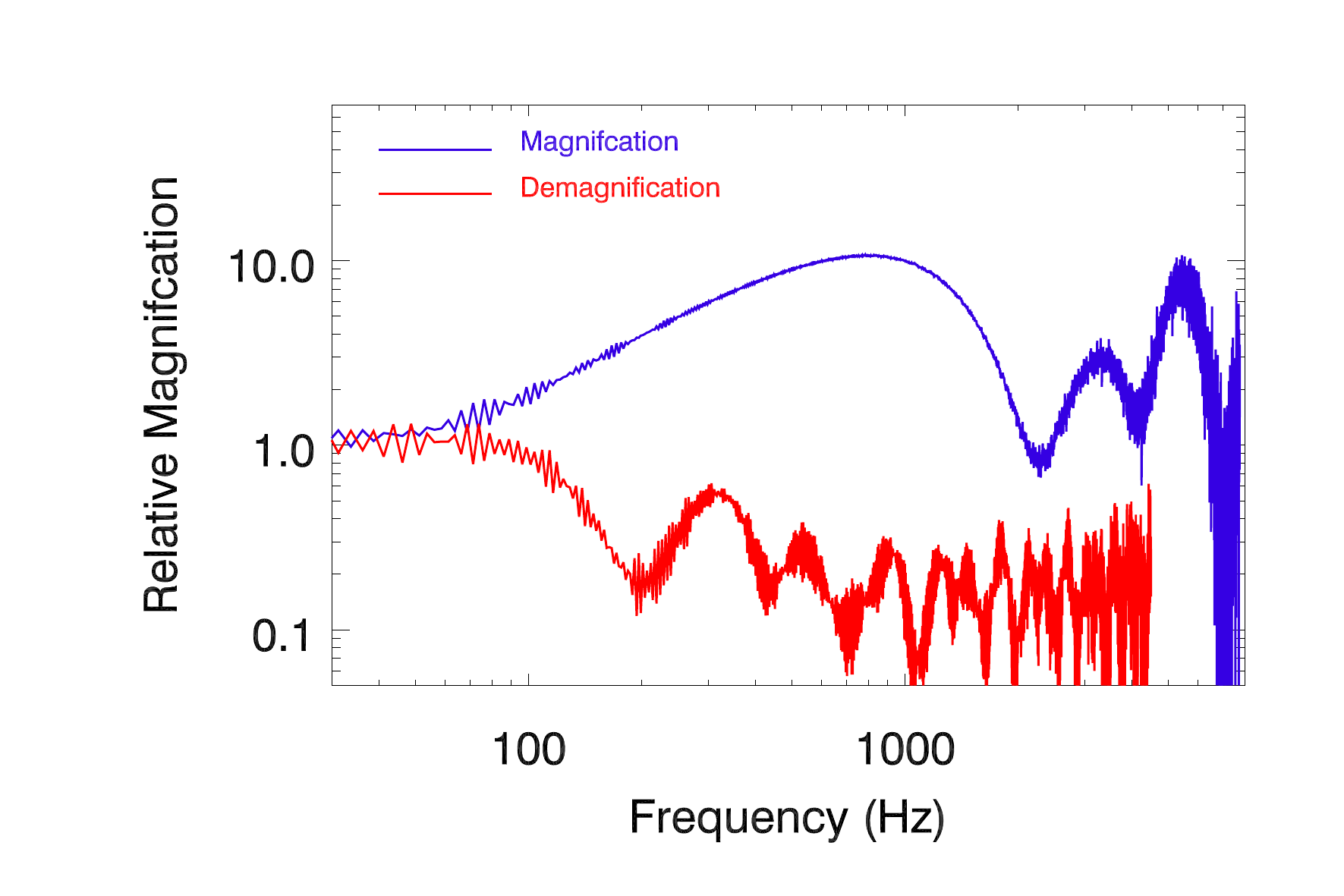}
      \caption{Magnification for the two sources shown in the right panel of Fig.~\ref{Fig_Arcs_TimeDelay_POS_NEG_3A} as a function of frequency. The dark blue curve is for the white source and the red curve is for the yellow source. 
              }
         \label{Fig_Fw_NEG_3A}
   \end{figure}

The corresponding magnification is shown in Fig.~\ref{Fig_Fw_POS_3A}. The color coding is the same as in Fig.~\ref{Fig_Ft_POS_3A}. For the white source in  Fig.~\ref{Fig_Arcs_TimeDelay_POS_NEG_3A} (dark-blue curve), a significant fluctuation is observed at $\approx 200$ Hz. Above this frequency, the magnification increases quickly up to factor $\approx 4$ with respect to the macromodel magnification at $\approx 300$ Hz to later decline again. At large frequencies, the magnification converges to the geometric optics limit, that is $\approx 1.5$ times the macromodel value. The red curve shows a similar, but milder, variability at LIGO/Virgo frequencies. 
In the light blue curve, the larger time delay imprints fluctuations in the magnification at even lower frequencies but these fluctuations are smaller in magnitude. It is important to note that, in the last two examples, even though at high frequencies the magnification converges to the geometric optics limit (below the macromodel value), and at very low frequencies the magnification converges to the macromodel value, in the intermediate ($\sim 100$ Hz) frequency range (the one best probed by LIGO/Virgo) the magnification can exceed that of the macromodel value improving the detectability of these events.

It is important to note that, although the geometric time delay (i.e, the separation between microimages) is significantly larger in the case shown in Fig.~\ref{Fig_Arcs_TimeDelay_POS_NEG_3A} than the corresponding case shown in the left panel of Fig.~\ref{Fig_CC_Caustics_Case2}, the time delays are still comparable. This is because the gravitational time delay is smaller when the macromodel potential is shallower (i.e at larger macromodel magnifications), but also the potential of the microlenses is smaller. 

\subsection{Macroimages with negative parity}

   \begin{figure*}
   \centering
   \includegraphics[width=18.0cm]{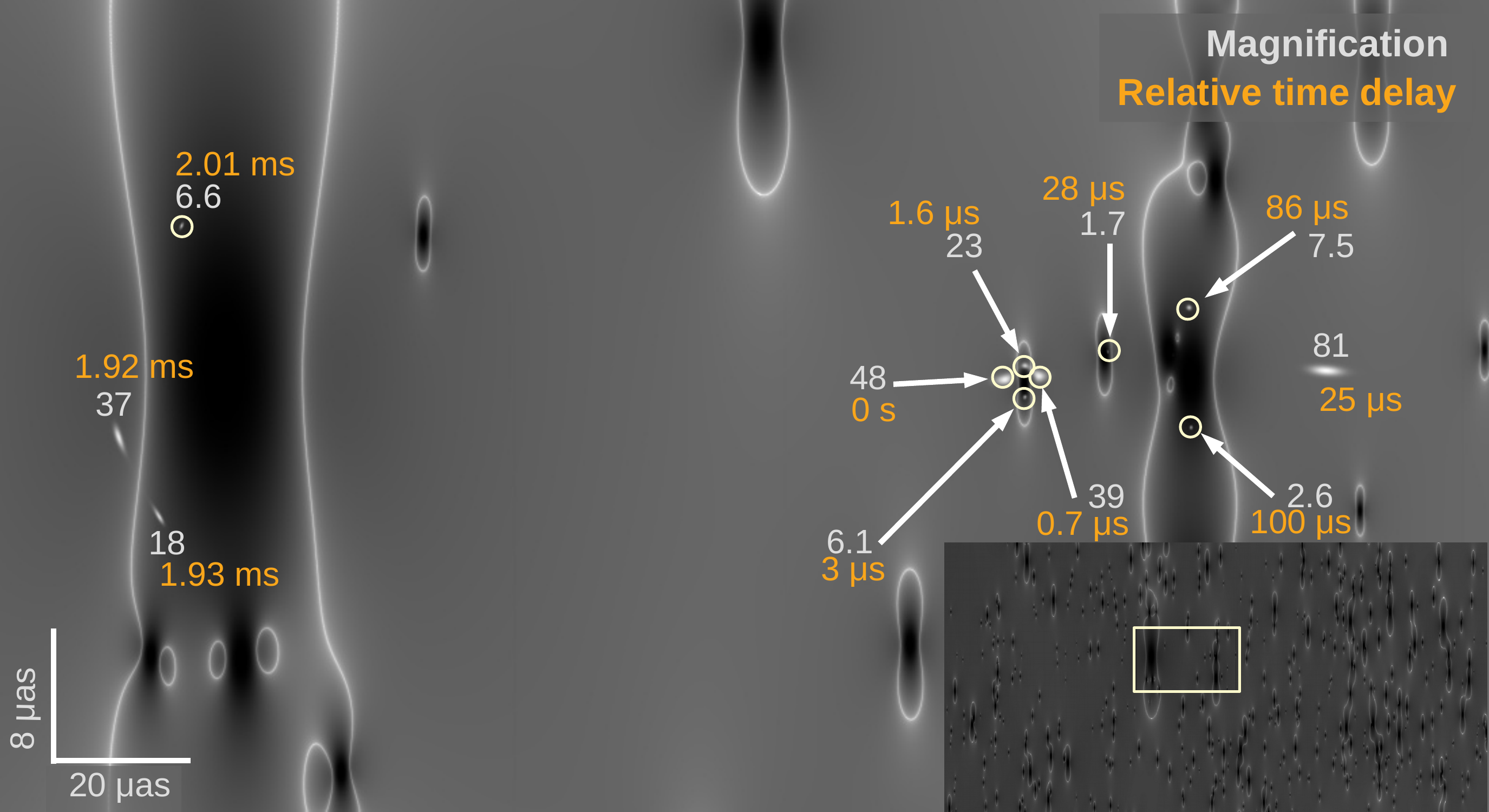}
      \caption{Zoom-in the region around a pair of microlenses with 11.3 \Msun and 2.2 \Msun in our simulated map. The entire simulated region is shown in the bottom right corner with the zoomed region marked with a rectangle. The arcs formed by a background source are shown in white. Small microimages are marked with a white circle.  
      The magnifications and relative time delays are shown for each microimage. 
              }
         \label{Fig_Crowded_Magnif}
   \end{figure*}
We place the same two sources at the same separation as in the previous subsection but the macromodel magnification is changed to its opposite value, that is $\mu_{macro}=-150$. 
When the microlens is on the side with macromodel negative parity, we find again the usual gap of low magnification between caustics (see small figure in the right panel of Fig.~\ref{Fig_Arcs_TimeDelay_POS_NEG_3A}). In the configuration of the example shown in the figure, the caustic regions overlap, so a source placed in that overlapping region produces multiple images around the critical curves of both microlenses as shown in the figure. If the caustics do not overlap, but instead one of the microcaustics falls inside the low magnification region of the other microlens, and a source is placed within that caustic region, images with large amplification factors would form only around one of the microlenses while the other microlens would form only microimages with relatively small magnification factors. Sources placed in the low magnification regions would form two microimages if the source is placed in only one of the two low magnification regions or four microimages if the source is placed in the overlapping low magnification region. In Fig.~\ref{Fig_Arcs_TimeDelay_POS_NEG_3A}) we show two sources, one white and one red. The white one is placed in the overlapping region of two caustics while the red one is placed in the low magnification region of one of the microlenses. The white source produces microimages with relatively moderate time delays ($\approx 1.5$ millisecond) between the microimages with the largest magnification. Longer time delays are only observed at the microimages with the smallest magnification. The red source (for which we do not show the counterimages) produces two small microimages near one of the microlenses.

The observed magnification of the GW of these two sources are shown in Fig.~\ref{Fig_Fw_NEG_3A}. The blue curve is for the white source and the red curve is for the red source. In the blue curve, we observe that, at low frequencies, the GW is insensitive to the microlens doublet as expected. However, within the frequency range proved by LIGO, the magnification increases by almost an order of magnitude which should imprint a modulation in the observed strain of a factor $\approx 3$ between the early and late periods of the GW signal. The red curve shows the case of the source in the low magnification region. Between 100 Hz and 200 Hz the GW is demagnified by as much as a factor 5. Between 100 Hz and 300 Hz, it is re-amplified by a factor almost 3 and above 300 Hz is demagnified again up to 400 Hz and so on. This very distinctive behaviour could be used to identify these events. Ironically, the GW that is less magnified may be easier to detect, since at low frequencies it is still being magnified by the macromodel, but at higher frequencies the GW would fade and oscillate.

\section{Field with stellar masses}\label{sect_crowded}\label{sect_ResultsIV}
In the previous sections we have studied simple, but pedagogical and useful, models with one massive microlens (100 \Msun) or a pair of intermediate mass microlenses ($\approx 30$ \Msun) in macromodel potentials with magnifications of $\approx 30$ and $\approx 150$ respectively. These previous examples are useful to understand the effect from single sources, or pairs of sources such as PBHs, but may not represent the most likely scenario. Instead, in real observations, most of the microlenses will be of stellar origin. Their abundance can be estimated based on the observed surface brightness that can be modelled with SEDs and a given IMF. In this section we address the question of the type of signal that can be produced when the GW travels through a field populated with stellar microlenses, and near a region of high magnification. 

For typical situations near critical curves, one finds that the surface mass density of microlenses contributing to the intracluster medium, or in the outskirts of galactic lenses, is in the range of a few to a few tens of solar masses per parsec$^2$. When fitting the observed flux, depending on whether one assumes a bottom heavy IMF such as a Salpeter IMF or a shallower IMF in the low-mass end such as a Chabrier IMF, the difference in the estimated surface mass density of microlenses can be typically a factor $\approx 2$ \citep{Morishita2017}. 

When a GW is observed with large magnification (due to the a macromodel), two macroimages are usually produced by the macromodel (galaxy or cluster) near one of the critical curves of the galaxy or cluster. At short distances from the critical curve, the role of microlenses from the galaxy halo, or intracluster medium, becomes important when the background source is small. This regime has been studied in detail in earlier work in the context of lensing of distant luminous stars \citep{Kelly2018,Diego2018a,Venumadhav2017}. For GWs there is no previous study (to the best of our knowledge) on the effect of a population of microlenses on a GW that is magnified by a large factor.  

   \begin{figure}
   \centering
   \includegraphics[width=9.0cm]{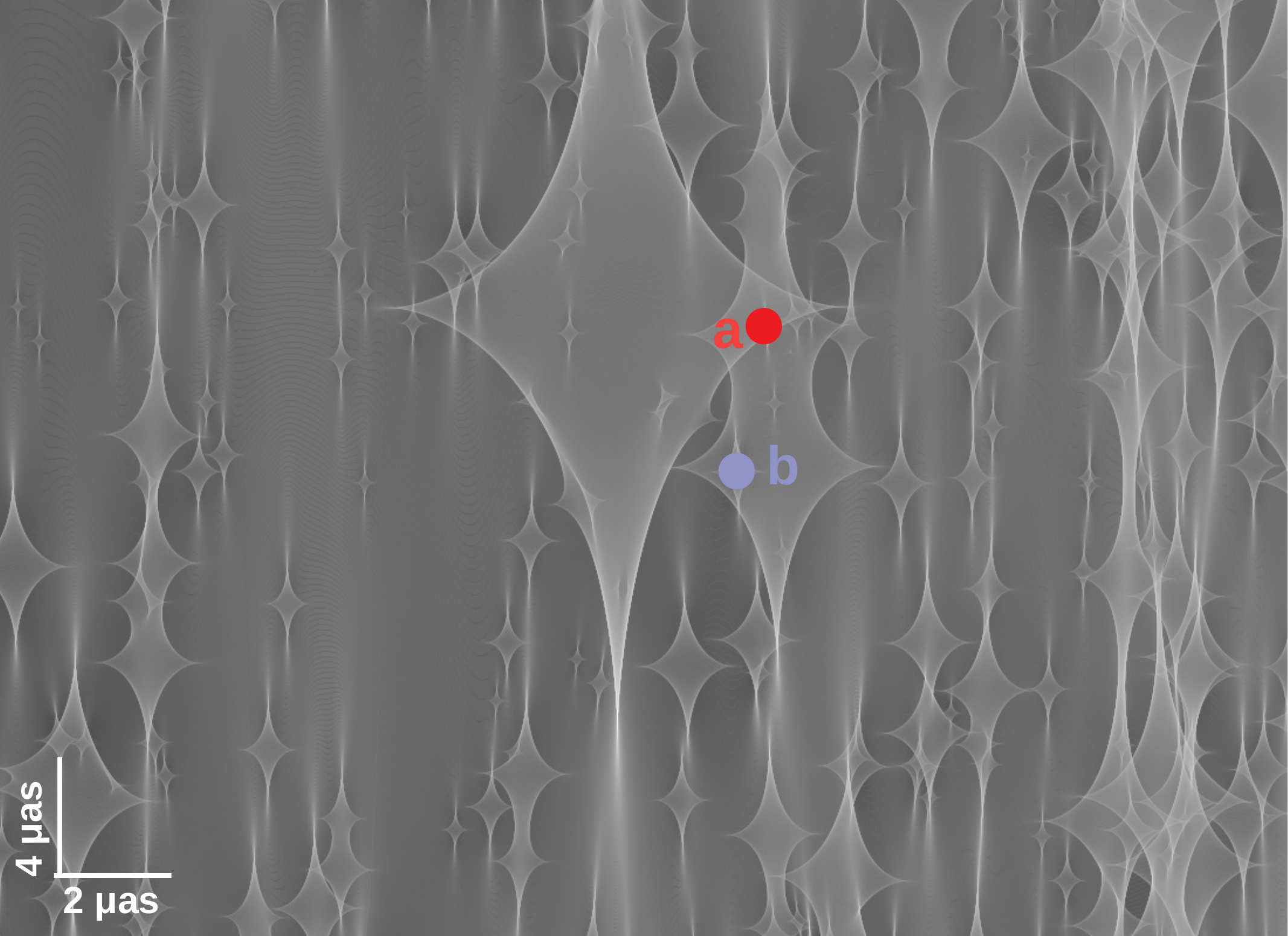}
      \caption{Caustic region on the side with positive parity. This region shows approximately half the simulated region. The largest microcaustic corresponds to the largest microlens of 11.3 \Msun. The red dot marks the position of the background source for which the microimages are shown in  Fig.~\ref{Fig_Crowded_Magnif}.  
              }
         \label{Fig_Crowded_Caust_POS}
   \end{figure}

We simulate an area in the image plane $1.2 {\rm mas}\times 1.2 {\rm mas} = 1.44 {\rm mas}^2$ with a surface mass density of microlenses of $\approx 12$ \Msun per parsec$^2$. This is a typical value found in real situations near critical curves of macromodels where the magnifications can be large \citep[this is for instance one of the values used to interpret the Icarus event][]{Kelly2018}. In general terms, we follow the procedure described in \cite{Diego2018b}. The full simulation box has a size of $1.44 {\rm mas}^2 = 37.3 {\rm pc}^2$ (at $z_s=2$), but we compute the deflection field and potential only in the central rectangular region of  $1.2\times 0.2$ mas$^2$) (or target region) with a resolution of 20 nanoarcsec per pixel.
In the original $1.44^2$ arcmin$^2$ area, we distribute randomly 2558 microlenses drawn from the stellar and remnant mass distribution computed for a Chabrier IMF to 0.01 Msun and the \cite{Fryer2012} initial-final mass function (with $z=0.5$). The oldest living star has an initial mass of 1.5 Msun, and we use the mass-dependent binary fraction from \cite{Duchene2013}. 
The total stellar mass in the simulated region is 430 \Msun  which results in $11.7$ \Msun/pc$^2$. The most massive microlens (star or remnant) in our simulation has a mass of 19.3 \Msun, although it falls outside the target region. Within the target region, the most massive microlens has a mass of 11.3 \Msun.  Our target region is large enough to be considered representative of the underlying distribution of microlenses. 
The parameters of the macromodel are given by the two values of the magnification, $\mu _r$ and $\mu _t$. These values determine the magnitude of the convergence, $\kappa$, and shear, $\gamma$. From $\kappa$, and $\gamma$ we derive the deflection field and potential of the macromodel following the procedure described in section 2. 

\subsection{Positive parity}
As before, we consider the scenarios of positive and negative macromodel parities separately. For a macromodel with $\mu _r=50$ and $\mu _t=3$ the resulting magnification in the lens plane is shown in Fig.~\ref{Fig_Crowded_Magnif}  (on the side of the lens with positive parity). The critical regions correspond to the white curves. The smaller picture on the bottom-right side is for the entire target region, while the enlarged picture is a zoom of the rectangular region highlighted in the small picture. The large microlens on the left side has 11.3 \Msun. 
The caustics in the source plane are shown in Fig.\ref{Fig_Crowded_Caust_POS}. The large microcaustic near the centre corresponds to the 11.3 \Msun microlens.

   \begin{figure}
   \centering
   \includegraphics[width=9.0cm]{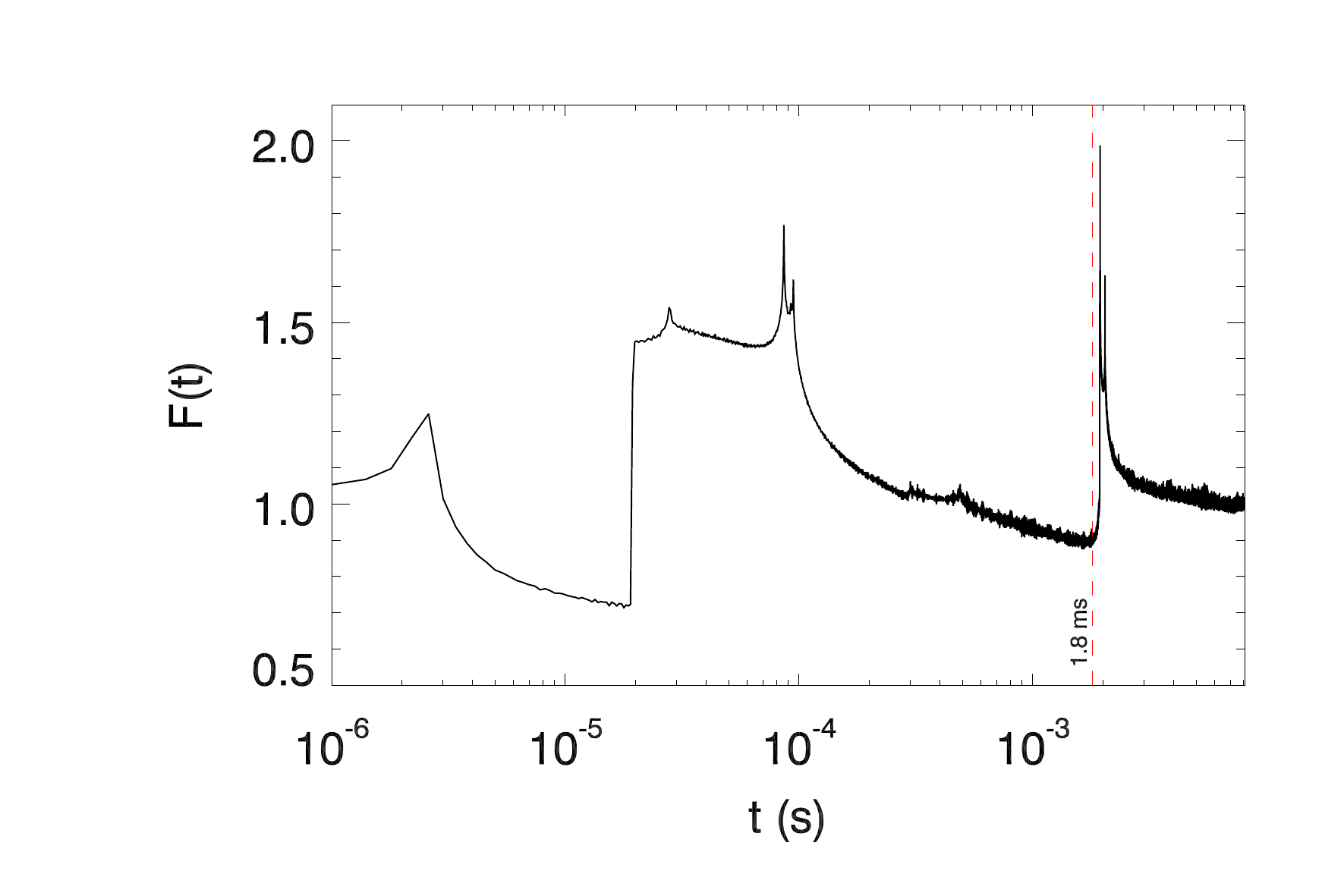}
      \caption{$F(t)$ for the red source shown in Fig.\ref{Fig_Crowded_Magnif}. Below 1 microsecond there is another peak not shown in this plot. The vertical dashed line marks the time delay of 1.8 ms.  
              }
         \label{Fig_Crowded_Magnif_Ft}
   \end{figure}

   \begin{figure}
   \centering
   \includegraphics[width=9.0cm]{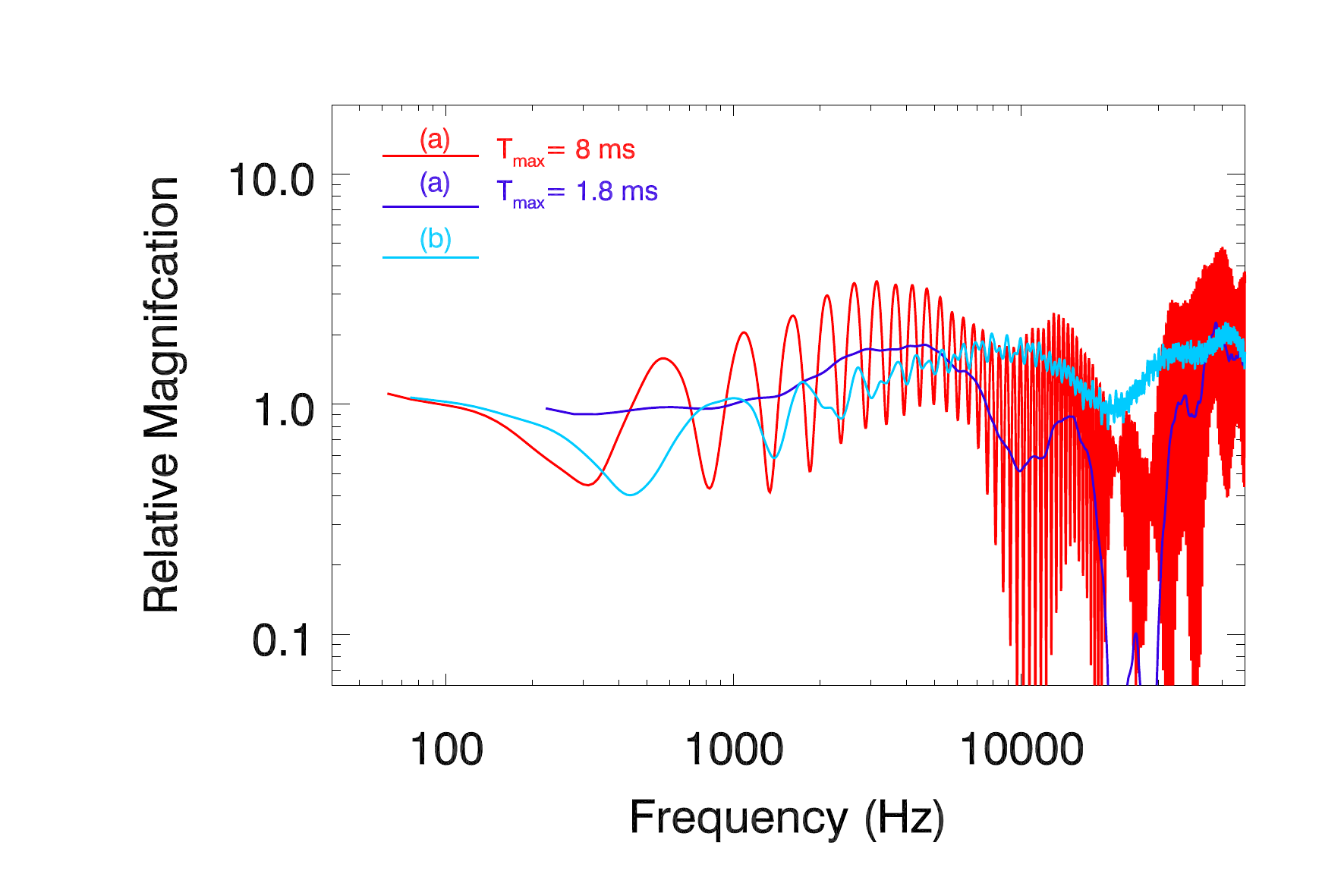}
      \caption{$F(w)$ for the two lensed sources shown in Fig.~\ref{Fig_Crowded_Magnif} (a and b). The red and dark blue curves are for source a. The dark blue curve is the corresponding amplification that it is obtained if only the peaks below 1.8 ms are considered.  The light blue curve is for the source b in Fig.~\ref{Fig_Crowded_Magnif}.
              }
         \label{Fig_Crowded_Magnif_Fw_POS}
   \end{figure}

For the models considered in this work, the number of microlenses (stars and remnants) above $\approx 10$ \Msun is scarce. The time delays between microimages are typically $\sim 0.1$ ms producing little noticeable effects at LIGO/Virgo frequencies. A few exceptions, where two microcaustics overlap, may result in larger time delays but usually still below 1 ms because small microlenses still need to be physically close to each other, in order for their small microcaustics to overlap. The probability of overlapping caustics is largest around the caustics of the most massive microlenses (black holes with >5 \Msun) since these caustics cover a relatively wide area in the source plane.  For instance, in Fig.\ref{Fig_Crowded_Caust_POS}, a source is placed inside the caustic of the largest microlens in our target region (with 11.3 \Msun). The source is also simultaneously inside the microcaustics of other smaller microlenses. The resulting microimages are shown in Fig.~\ref{Fig_Crowded_Magnif} in white. Next to each microimage we indicate their individual magnifications and time delay relative to the microimage that arrives first.  Since the time delays between microimages with high magnification  can be several ms, we expect the signature of interference to manifest itself at LIGO/Virgo frequencies. Note, however, how the group of microimages on the right side has relative times delays smaller than 0.1 ms. If the large microlens on the left were not present, these microimages would not result in significant distortions to the magnification at LIGO/Virgo frequencies.  

The corresponding $F(t)$ curve, for the configuration shown in Fig.~\ref{Fig_Crowded_Magnif} is shown in Fig.~\ref{Fig_Crowded_Magnif_Ft}. The small time delays between microimages produced by the smaller microlenses are responsible for the multiple peaks below 0.1 ms (below 1 microarcsec there is another small peak not shown in the figure). The three microimages around the large microlens of 11.3 \Msun produce the discontinuity at around 2 ms, and two small, almost overlapping logarithmic peaks.

In order to show what would have been the effect, should the large  11.3 \Msun microlens not be present, we consider also the \Ft up to 1.8 ms (marked with a vertical dashed line in the figure) and compute the amplification in both scenarios. 
Figure~\ref{Fig_Crowded_Magnif_Fw_POS} shows the relative magnification when considering the full spectrum ($t_{max}=8$ ms, red solid curve) and when considering only the spectrum up to 1.8 ,ms (blue curve). In general, the red curve traces the blue curve, but has the additional harmonic produced by the peak in \Ft at $\approx 2$ ms.  The overlap of caustics from the large microlens and the smaller ones results in a modulation of the magnification at LIGO frequencies, where at frequencies $\approx 300$ Hz the magnification is a factor $\approx 2$ times smaller than at frequencies $< 100$ Hz (red curve). If the large microlens was not present, the overlapping of the two smaller caustics is not be sufficient to produce significant effects at LIGO/Virgo frequencies (blue curve).

\subsection{Negative parity}
\begin{figure}
   \centering
   \includegraphics[width=9.0cm]{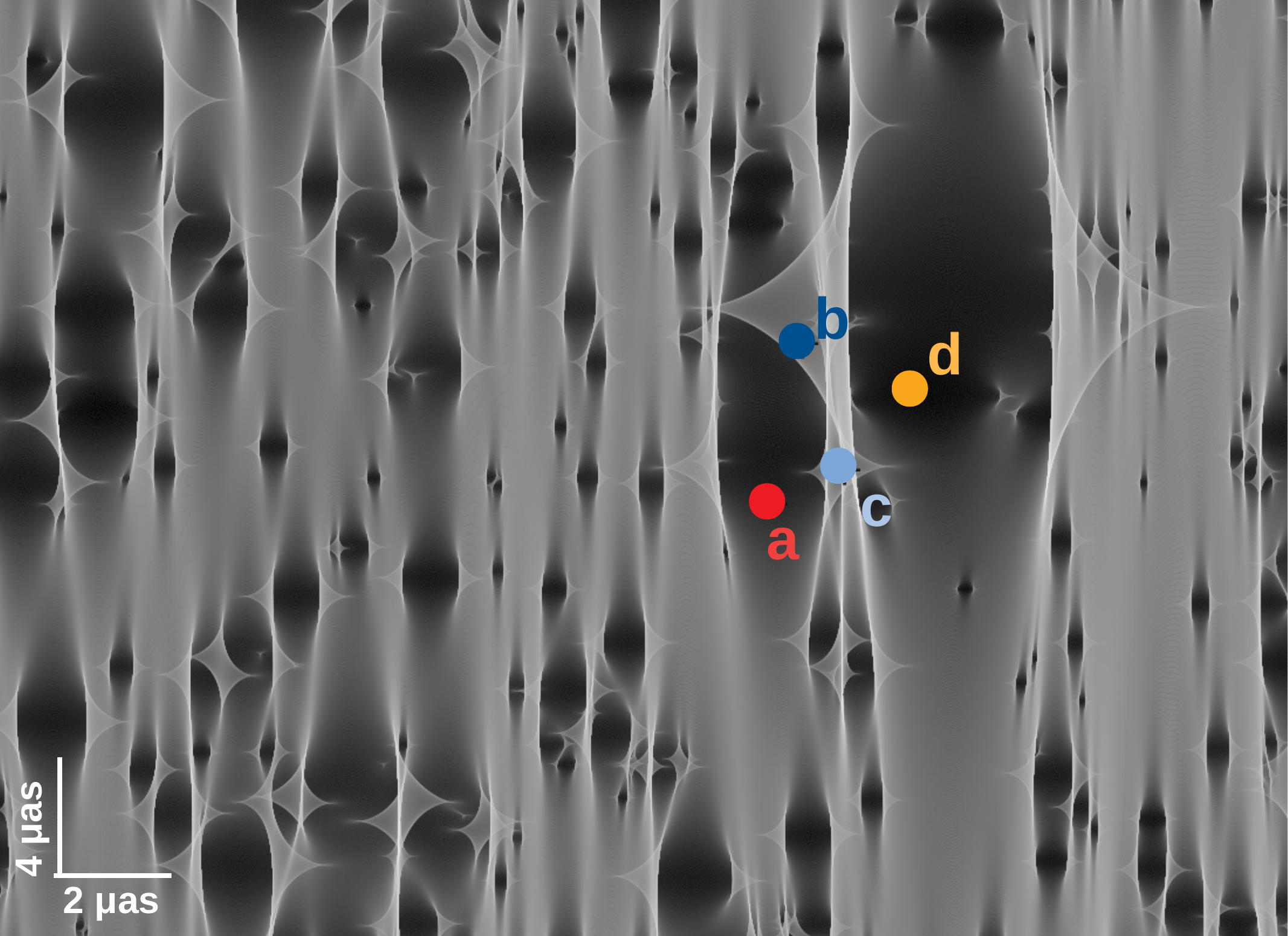}
      \caption{Caustic region on the side with negative parity. Four source positions are labled with a, b, c and d.
              }
         \label{Fig_Crowded_Caust_NEG}
   \end{figure}

\begin{figure}
   \centering
   \includegraphics[width=9.0cm]{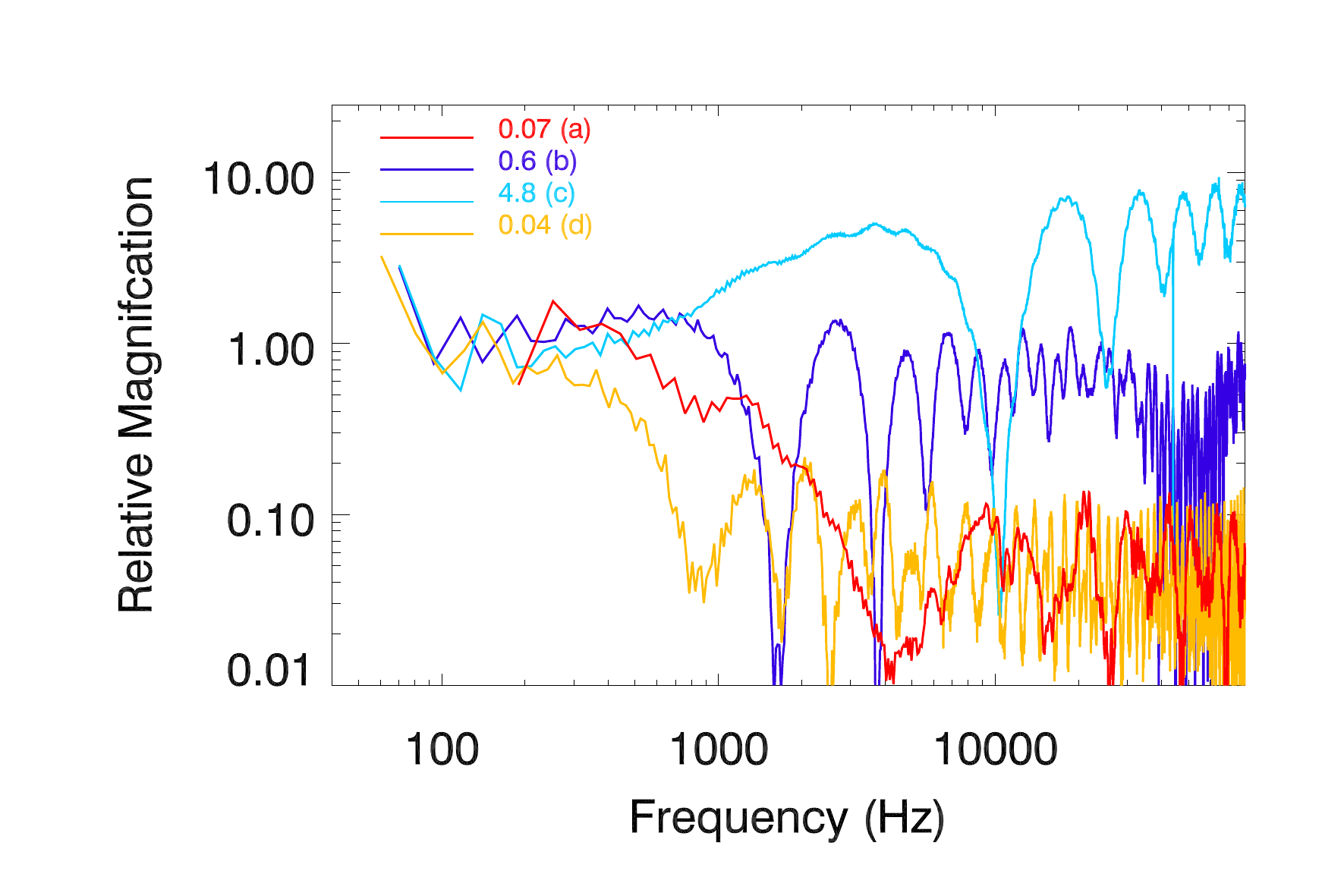}
      \caption{$F(w)$ for the lensed sources shown in Fig.\ref{Fig_Crowded_Caust_NEG}. The color of the curves is the same as the color of the sources in Fig.\ref{Fig_Crowded_Caust_NEG}. 
              }
         \label{Fig_Crowded_Magnif_Fw_NEG}
   \end{figure}

When the parity of the macromodel is negative, we have seen in previous sections that microcaustics contain a relatively large area of low magnification in between the regions of high magnification. Figure \ref{Fig_Crowded_Caust_NEG} shows the caustics in a fraction of the simulated source plane. The distribution of microlenses is the same as the one discussed in the previous subsection. The large microlens in the image is also the same 11.3 \Msun microlens discussed above. We place sources in three positions marked with (a), (b) and (c). The magnifications as a function of GW frequency are shown in Fig.~\ref{Fig_Crowded_Magnif_Fw_NEG}. Sources (b) and (c) show the smallest departure from unity in the relative magnification at LIGO frequencies. Source (a) shows a deviation from unity, only at the largest LIGO frequencies that could be observed. Note how source (a) is not placed in the largest microlens but on a more moderate (and common) one with mass $\approx$ 2 \Msun. On the side with negative parity, low magnification is more likely than high magnification so it is more likely than a macroimage on the side with negative parity suffers a larger distortion than a macroimage on the side with positive parity. If microlenses with masses as low a 2 \Msun can produce interference patterns at LIGO/Virgo frequencies, this may affect the detectability of counterimages with negative parity as templated searches without lensing considerations could miss these patterns.


\section{Effect over the GW strain}\label{sect_strain}
\begin{figure}
   \centering
   \includegraphics[width=9.0cm]{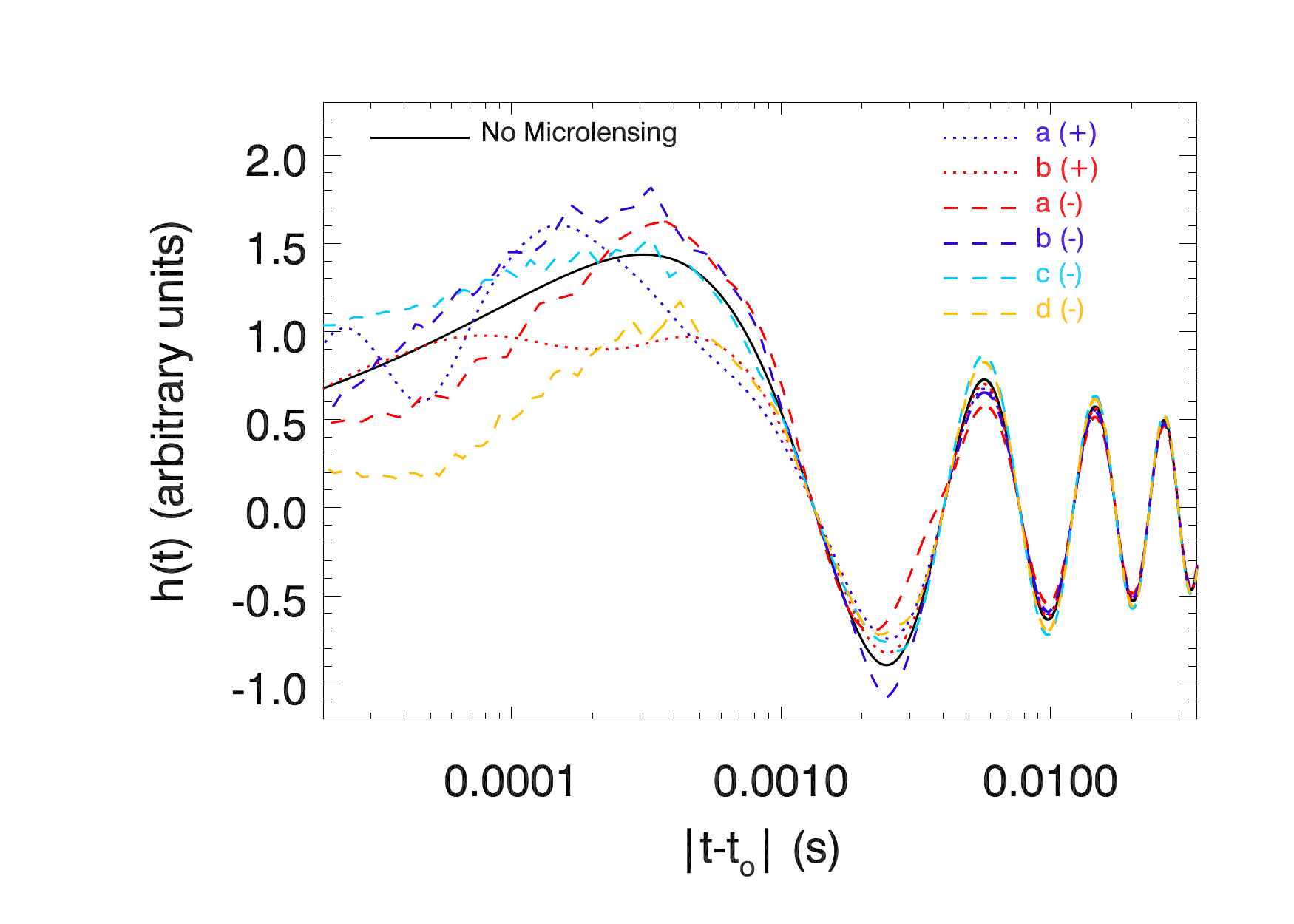}
      \caption{Simulated strain with no-lensing of the coalescence of a BBH with 10 and 12 \Msun\ at z=2 (solid black line). Note that in order to zoom in the highest frequencies, we plot the strain as a function of $|t-t_o|$ where the time, $t_o$, corresponds to zero in the x-axis. The dashed lines show the distortion of the strain for the different microlensing scenarios (by stellar bodies) discussed in Figures 15 and 17.  
              }
         \label{Fig_Strain}
   \end{figure}

The results in the previous section show the effect of microlensing over the magnification of GWs. A more direct visualization of the real effect can be appreciated when the magnification is applied over a simulated strain. In order to do this, we assume a merging event of a BBH at z=2 with $m_1=m_2=10$ \Msun\ amplified by some larger macromodel magnification, $\mu \approx 150$ (as in the previous section). The unperturbed strain from the last orbits of the merging event is shown in Fig.~\ref{Fig_Strain} as a black solid line. The units in the y-axis are arbitrary and incorporate already the large magnification factor from the macromodel, $\sqrt{\mu}$. The dotted and dashed lines show the distorted strain as a consequence of the interference effect and the change in magnification as a function of frequency. The cases considered in this plot are the same shown in Figures 15 and 17 above. In particular, dotted lines are for the sources "a" and "b" in the side with positive macromdoel parity while dashed lines are for the sources "a", "b", "c", and "d" in the side with negative macromdoel parity.  
In general, for microlenses with stellar masses as those considered in the previous section the effect is small and it is concentrated in the last cycles of the merger. Unsurprisingly, the case imposing the largest effect is the orange dashed line which corresponds to orange source "d" in Figure 16 (orange curve in Figure 17). The microlens that causes this distortions is the most massive in our simulation (11.3 \Msun). However, even sources placed in the caustic region of a smaller microlens, like source "b" in Fig.~\ref{Fig_Crowded_Caust_POS} (in the caustic region of a microlens with a mass of 2.2 \Msun) still show significant departures in the amplitude of the last cycle. 

Usually, one would quantify the detectability of these lensed events by Bayesian hypothesis testing using templated searches as in~\cite{Lai2018}. 
However, as no generic waveform templates exist for these microlensed waveforms, quantifying the detectability of these events in this way is difficult. 

\begin{figure}
   \centering
   \includegraphics[width=9.0cm]{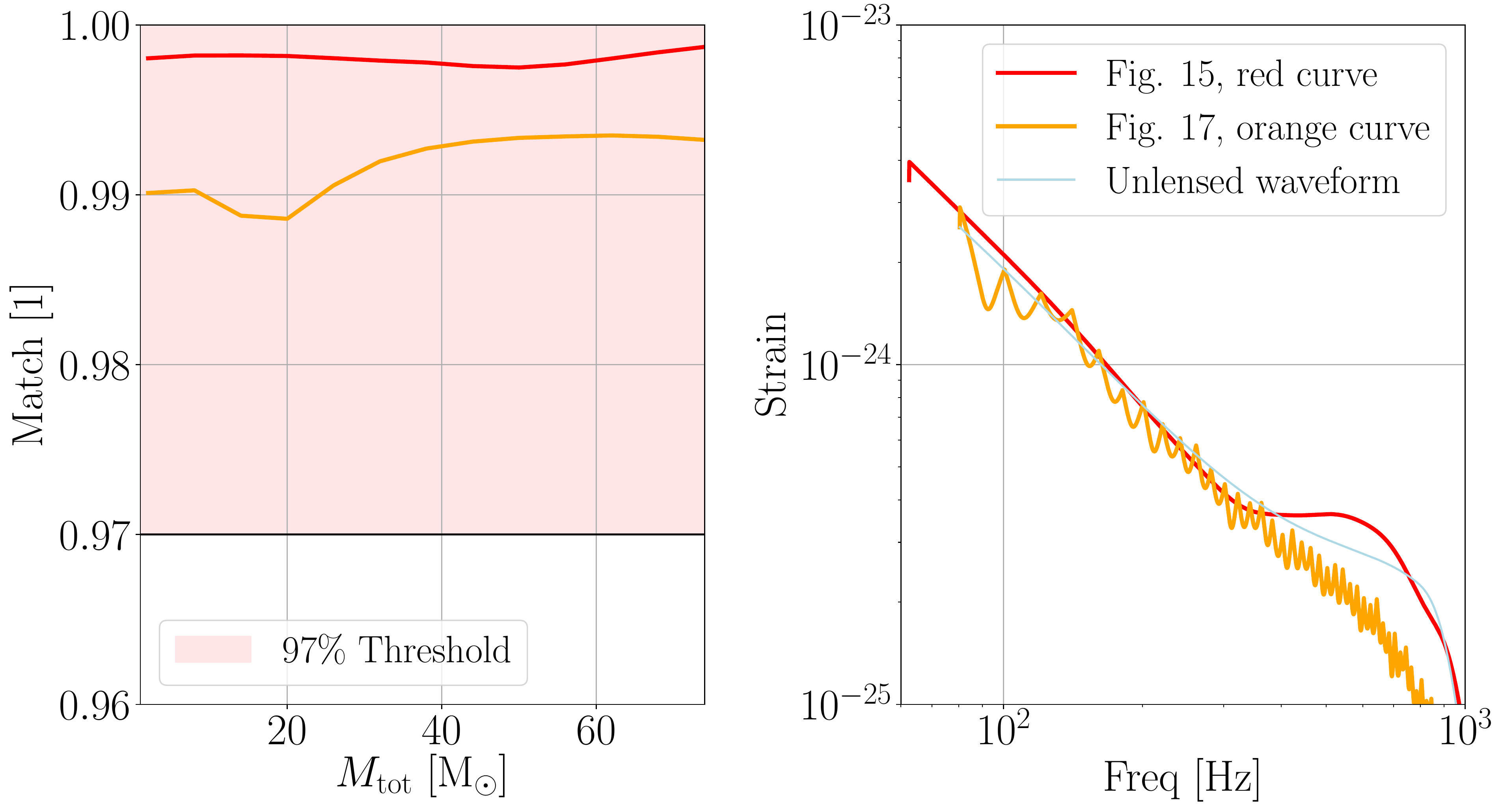}
      \caption{{\it Left panel:} Waveform match between the microlensed and lensed waveform as a function of the total mass of the binary $M_{\rm tot}$. 
      {\it Right panel:} The strain amplitude as a function of frequency for the lensed and unlensed waveform with $M_{\rm tot}=20$ \Msun. 
      The red and orange curves utilize the lens configurations and magnification factors shown in Figs.~\ref{Fig_Crowded_Magnif_Fw_POS} and~\ref{Fig_Crowded_Magnif_Fw_NEG}, respectively, while the black line (right panel) shows the unlensed waveform. 
      The waveform corrections are at percent level. 
              }
         \label{Fig_StrainMatch}
   \end{figure}

We instead quantify the correction to the waveform introduced by lensing by comparing the waveform match $m(h_{\rm lensed}, h_{\rm unlensed})$ (the normalized inner product~\cite[e.g.][]{maggiore}) between the lensed and unlensed versions, weighted with the Advanced LIGO power spectral density at design sensitivity. 
Corrections of a few $\sim 3$ \%, which are roughly equivalent to corrections introduced by spin effects, could have hope of leaving discernible patterns in the LIGO/Virgo strains. 
We generate two representative examples of lensed waveforms for the lens setting shown in Figs.~\ref{Fig_Crowded_Magnif_Fw_POS} (red curve) and~\ref{Fig_Crowded_Magnif_Fw_NEG} (orange curve), and compute the match between their unlensed counterparts. 
Interesting modulation, or "beating patterns", are present in the waveforms (Fig.~\ref{Fig_StrainMatch}, right panel) in comparison to the unlensed waveform, that may be discernible in the future. Note that in the orange curve, the small scale fluctuations are artifacts due to the noise in the computation of the relative magnification. 
The reason why the strain amplitude is larger at lower frequency is due to the binary spending longer time at these frequencies. Alternatively, the y-axis in the right panel of Fig.~\ref{Fig_StrainMatch} can be also interpreted as the power spectrum of the strain as a function of time. 
However, the correction introduced by lensing in the waveforms is at around \% level (Fig.~\ref{Fig_StrainMatch}, left panel), which could be difficult to detect at current sensitivities within LIGO/Virgo due to parameter degeneracies, noise and the Occam's razor (which at similar performance, favours models with fewer parameters). 

One difficulty in detecting these small modulations induced by microlensing is that they mainly introduce amplitude modulations in the waveforms that are hard to discern with matched filtering schemes. 
This is in contrast to phase modulations which could be more suitable for the current search methods. 
Another difficulty is that the most interesting lensing features range fall just slightly off the target frequency band of the LIGO/Virgo detectors.
While the few microlensing examples we have shown introduced relatively small corrections to the waveform, certain microlensing effects could perhaps be visible at design sensitivities. 
For example, we have verified that time delays of order $\sim 3$ milliseconds from two micro-images could introduce waveform corrections at the 10\% level, assuming relative magnifications of 1 : 2. 
As the next generation of gravitational wave detectors are being built, the improved sensitivities they offer could open the observational window for these events. 
Improvements especially above 500 Hz, where the representative lens configurations here have shown interesting modulations in the amplification factors, would improve our odds of detecting these events. 
These detections could then be used to investigate and constrain populations of microlenses (including an alleged population of PBHs for which constraints could be imposed on their abundance through this technique), similarly to the studies in~\cite{Diego2018a}. 

Although not discussed in this paper, it may be important to consider the effect that undetected microlensing can have in the derivation of parameters, specially those that depend more on the last cycles, or the ringdwon phase  of the strain (highest frequencies), such as the spin. 


\section{Discussion and conclusions}\label{sect_concl}
GWs that are magnified (by a macromodel) by large factors can also be affected by the microlenses inside the macromodel. If the macromodel magnification is large enough (i.e, when the saturation regime mentioned in section \ref{sect_theoryI} is reached), microlensing is not only possible, but unavoidable. These microlenses imprint an interference pattern in the lensed GW, leaving a characteristic signature that can be used to unequivocally classify the event as a strongly lensed event. Hence, most if not all, GWs that are magnified by extreme factors are expected to exhibit some distortion in their strains as a consequence of the modulation of the magnification as a function of frequency. 

Microlenses near a macromodel critical curve behave as microlenses with a larger effective Einstein radius, and consequently the geometric time delay increases accordingly (but so does the effective lensing potential, compensating partially the effect). Time delays between two microimages of the GW of order 1 ms result in interference patterns of the GW that affect the strain at LIGO/Virgo frequencies. We have shown how microlenses with masses of order 100 \Msun in regions of the lens plane where the macromodel magnification is of order a few tens imprint very distinctive patterns in the magnification at LIGO/Virgo frequencies. For macroimages of negative parity, the role of microlenses becomes more important. In this case, the likelihood of a source falling in a region of low magnification is significantly higher than it is for high magnification when macroimages have positive parity. 
Morever, the more likely low magnification regions in the source plane imprint larger time delays in the observer plane, making interference of macroimages with negative parity, not only more likely, but also more pronounced at LIGO/Virgo frequencies. 

Another important consequence of large macromodel magnifications is that the probability of microcaustic overlapping grows with the macromodel magnification. Two relatively small microcaustics can conspire together to produce a larger time delay if their microcaustics are overlapping. We demonstrate this effect with microlenses having a few tens of \Msun but also with realistic distributions of stellar bodies and their remnants. In the later case, we find that microlenses with masses less than a few \Msun are less likely to produce significant distortions in the magnification pattern at LIGO/Virgo frequencies, even if their microcaustics are overlapping. However, overlapping microlenses with masses of a few solar masses, and close to the critical curve of the macromodel can result in time delays of order one millisecond, or larger, hence producing significant distortions in the magnification at LIGO/Virgo frequencies.  
For macroimages with negative parity, even a modest mass of $\approx 2$ \Msun can produce deviations at the 20\% level in the highest frequency range probed by LIGO/Virgo ($\approx 500$ Hz) which could have an impact in the detectability of the signal, or at the parameter estimation level. 
Macroimages with negative parity are more likely to produce pronounced effects within LIGO/Virgo than macroimages with positive parity. Howevere, the larger probability of de-magnification (relative to the magnification of the macromodel), specially at frequencies above 100 Hz, may represent a greater challenge to identify them in the noisy data. Consequently, strongly lensed GWs with negative  parity are more likely to be missed than their counterimages with positive parity. 

As already pointed out by earlier work, in the regime where wave optics is important, the effect of microlenses over a GW extend farther away from the caustic region. This increases the probability (or effective optical depth) of detecting a microlensing event through its distortion in the magnification. We have shown how interference can take place even at low magnifications (which would pass unrecognized as a lensed event). 
An appropriate optimal filter could perhaps unveil GW that present such interferences in certain lensing configurations. Detecting the interference signature in present data will be challenging given the relatively high noise level, parameter degeneracies, and Occam's razor argument. 

We have quantified this effect by estimating the percentage of correction introduced to the waveform by these lensing interference patterns. 
Since most of the effect is produced at frequencies above 100 Hz (for normal stellar mass microlenses), the most intriguing range falls just slightly off the optimal frequency range of the LIGO/Virgo detectors.
However, when future detectors such as the Einstein Telescope and Cosmic Explorer~\citep[see][] {3gdetectors1,3gdetectors2,3gdetectors3,3gdetectors4} become operational with much better sensitivities, detecting these events will become significantly more likely. 

An interesting application of lensing of GWs is using the time delay in the context of multimessenger astrophysics. As suggested by \cite{Baker2017}, these time delays induced by gravitational lensing of gravitational waves in unison with their electromagnetic counterpart could be used to constrain the cosmological models (for instance by setting bounds on the total neutrino mass). The time delay induced by microlenses can be inferred from the modulation of the observed strain. This time delay will normally be in the range of a millisecond to a few milliseconds. Although \cite{Baker2017} considers neutrinos and GWs, EM counterparts can be used as well to derive cosmological information \citep{Liao2017,Yang2018,Li2018}. Fast Radio Bursts (FRBs) would make ideal electromagnetic (EM) counterparts since their duration is also comparable to the time delay (and time delays of order a ms can be measured from them). However, the origin of FRBs is still unclear and probably not linked to BBH mergers. The ideal EM pulse must be associated with the merger event (in order for the pulse to follow the exact same path as the GW) leaving the coalescence of binary neutron stars as the optimal candidates. Future GW detectors, with sufficient sensitivity to observe lensed binary neutron star mergers, such as the Einstein telescope, may be able to study them. Also, in BNS or BNBH mergers, the chirp mass is smaller and consequently the frequency larger. For these type or mergers, microlensing events from stellar masses will be more relevant due to the higher frequency of the GW.
Observing a BNS through lensing requires large magnfication factors since the chirp mass (and consequently the signal-to-nose ratio) for these type of events is smaller and the optical depth for events at redshift below $\approx 0.3$ is also very small. In order to compensate for the increased luminosity distance (required to incrase the optical depth of lensing and see a lensed BNS), the magnifcation factors need to be large (hundreds or thousands). If BNS mergers are more common than BBH mergers, strongly magnified BNS with magnifications factors of several hundreds could be soon observed and microlensing events affecting these GWs could be studied in more detail.  

Although not discussed in this paper (and with little prospects of being a viable type of observations in the near future), our results are relevant also for radio astronomy in the range of the kHz and operating from space. In this regime, radio pulses like those emitted by pulsars or FRB (assuming they still have sufficient power at these very low frequencies) could be lensed by microlenses of stellar, or substellar masses, and result in interference patters similar to those presented in this work. 

\begin{acknowledgements}
J.M.D. acknowledges the support of projects AYA2015-64508-P (MINECO/FEDER, UE), funded by the Ministerio de Economia y Competitividad. 
J.M.D. acknowledges the hospitality of the Physics Department at the University of Pennsylvania for hosting him during the preparation of this work. The authors thank Mark Trodden for useful comments. 
O.A.H. is supported by the Hong Kong Ph.D. Fellowship Scheme (HKPFS) issued by the Research Grants Council (RGC) of Hong Kong.
The work of O.A.H., K.K., and T.G.F.L. was partially supported by grants from the Research Grants Council of the Hong Kong (Project No. CUHK14310816, CUHK24304317 and/or CUHK 14306218) and the Direct Grant for Research from the Research Committee of the Chinese University of Hong Kong.
\end{acknowledgements}

\bibliographystyle{aa} 
\bibliography{MyBiblio} 

\begin{thebibliography}{69}
\expandafter\ifx\csname natexlab\endcsname\relax\def\natexlab#1{#1}\fi

\bibitem[{Abbott {et~al.}(2017)Abbott, Abbott, Abbott, Abernathy, Ackley,
  Adams, Addesso, Adhikari, Adya, Affeldt, {et~al.}}]{3gdetectors2}
Abbott, B.~P., Abbott, R., Abbott, T., {et~al.} 2017, Classical and Quantum
  Gravity, 34, 044001

\bibitem[{{Abbott} {et~al.}(2016){Abbott}, {Abbott}, {Abbott}, {Abernathy},
  {Acernese}, {Ackley}, {Adams}, {Adams}, {Addesso}, {Adhikari}, \&
  et~al.}]{2016PhRvD..93l2004A}
{Abbott}, B.~P., {Abbott}, R., {Abbott}, T.~D., {et~al.} 2016, \prd, 93, 122004

\bibitem[{{Abbott} {et~al.}(2017){Abbott}, {Abbott}, {Abbott}, {Abernathy},
  {Acernese}, {Ackley}, {Adams}, {Adams}, {Addesso}, {Adhikari}, \&
  et~al.}]{2017PhRvD..95d2003A}
{Abbott}, B.~P., {Abbott}, R., {Abbott}, T.~D., {et~al.} 2017, \prd, 95, 042003

\bibitem[{Abernathy {et~al.}(2011)Abernathy, Acernese, Ajith, Allen,
  Amaro-Seoane, {et~al.}}]{3gdetectors4}
Abernathy, M., Acernese, F., Ajith, P., {et~al.} 2011, available from European
  Gravitational Observatory, document number ET-0106A-10

\bibitem[{Akutsu {et~al.}(2018)Akutsu, Ando, Araki, Araya, Arima, Aritomi,
  Asada, Aso, Atsuta, Awai, {et~al.}}]{akutsu2018construction}
Akutsu, T., Ando, M., Araki, S., {et~al.} 2018, Progress of Theoretical and
  Experimental Physics, 2018, 013F01

\bibitem[{Aso {et~al.}(2013)Aso, Michimura, Somiya, Ando, Miyakawa, Sekiguchi,
  Tatsumi, Yamamoto, Collaboration, {et~al.}}]{aso2013interferometer}
Aso, Y., Michimura, Y., Somiya, K., {et~al.} 2013, Physical Review D, 88,
  043007

\bibitem[{{Baker} \& {Trodden}(2017)}]{Baker2017}
{Baker}, T. \& {Trodden}, M. 2017, \prd, 95, 063512

\bibitem[{{Broadhurst} {et~al.}(2018){Broadhurst}, {Diego}, \&
  {Smoot}}]{Broadhurst2018}
{Broadhurst}, T., {Diego}, J.~M., \& {Smoot}, III, G. 2018, arXiv e-prints
  [\eprint[arXiv]{1802.05273}]

\bibitem[{{Broadhurst} {et~al.}(2019){Broadhurst}, {Diego}, \&
  {Smoot}}]{Broadhurst2019}
{Broadhurst}, T., {Diego}, J.~M., \& {Smoot}, III, G.~F. 2019, arXiv e-prints
  [\eprint[arXiv]{1901.03190}]

\bibitem[{Cao {et~al.}(2014)Cao, Li, \& Wang}]{Cao2014}
Cao, Z., Li, L.-F., \& Wang, Y. 2014, Phys. Rev. D, 90, 062003

\bibitem[{{Carr} {et~al.}(2017){Carr}, {Raidal}, {Tenkanen}, {Vaskonen}, \&
  {Veerm{\"a}e}}]{Carr2017}
{Carr}, B., {Raidal}, M., {Tenkanen}, T., {Vaskonen}, V., \& {Veerm{\"a}e}, H.
  2017, \prd, 96, 023514

\bibitem[{{Chang} \& {Refsdal}(1979)}]{Chang1979}
{Chang}, K. \& {Refsdal}, S. 1979, \nat, 282, 561

\bibitem[{{Chang} \& {Refsdal}(1984)}]{Chang1984}
{Chang}, K. \& {Refsdal}, S. 1984, \aap, 132, 168

\bibitem[{{Chen} {et~al.}(2019){Chen}, {Kelly}, {Diego}, {Oguri}, {Williams},
  {Zitrin}, {Treu}, {Smith}, {Broadhurst}, {Kaiser}, {Foley}, {Filippenko},
  {Salo}, {Hjorth}, \& {Selsing}}]{Chen2019}
{Chen}, W., {Kelly}, P.~L., {Diego}, J.~M., {et~al.} 2019, arXiv e-prints
  [\eprint[arXiv]{1902.05510}]

\bibitem[{{Christian} {et~al.}(2018){Christian}, {Vitale}, \&
  {Loeb}}]{Christian2018}
{Christian}, P., {Vitale}, S., \& {Loeb}, A. 2018, \prd, 98, 103022

\bibitem[{{Dai} {et~al.}(2018){Dai}, {Li}, {Zackay}, {Mao}, \& {Lu}}]{Dai2018}
{Dai}, L., {Li}, S.-S., {Zackay}, B., {Mao}, S., \& {Lu}, Y. 2018, \prd, 98,
  104029

\bibitem[{Dai {et~al.}(2017)Dai, Venumadhav, \& Sigurdson}]{dai2017effect}
Dai, L., Venumadhav, T., \& Sigurdson, K. 2017, Physical Review D, 95, 044011

\bibitem[{{Dai} {et~al.}(2017){Dai}, {Venumadhav}, \& {Sigurdson}}]{Dai2017}
{Dai}, L., {Venumadhav}, T., \& {Sigurdson}, K. 2017, \prd, 95, 044011

\bibitem[{{Deguchi} \& {Watson}(1986)}]{Deguchi1986}
{Deguchi}, S. \& {Watson}, W.~D. 1986, \apj, 307, 30

\bibitem[{{Diego}(2018)}]{Diego2018b}
{Diego}, J.~M. 2018, arXiv e-prints [\eprint[arXiv]{1806.04668}]

\bibitem[{{Diego} {et~al.}(2018){Diego}, {Kaiser}, {Broadhurst}, {Kelly},
  {Rodney}, {Morishita}, {Oguri}, {Ross}, {Zitrin}, {Jauzac}, {Richard},
  {Williams}, {Vega-Ferrero}, {Frye}, \& {Filippenko}}]{Diego2018a}
{Diego}, J.~M., {Kaiser}, N., {Broadhurst}, T., {et~al.} 2018, \apj, 857, 25

\bibitem[{{Duch{\^e}ne} \& {Kraus}(2013)}]{Duchene2013}
{Duch{\^e}ne}, G. \& {Kraus}, A. 2013, \araa, 51, 269

\bibitem[{Dwyer {et~al.}(2015)Dwyer, Sigg, Ballmer, Barsotti, Mavalvala, \&
  Evans}]{3gdetectors3}
Dwyer, S., Sigg, D., Ballmer, S.~W., {et~al.} 2015, Physical Review D, 91,
  082001

\bibitem[{{Fryer} {et~al.}(2012){Fryer}, {Belczynski}, {Wiktorowicz},
  {Dominik}, {Kalogera}, \& {Holz}}]{Fryer2012}
{Fryer}, C.~L., {Belczynski}, K., {Wiktorowicz}, G., {et~al.} 2012, \apj, 749,
  91

\bibitem[{{Fukugita} {et~al.}(1992){Fukugita}, {Futamase}, {Kasai}, \&
  {Turner}}]{Fukugita1992}
{Fukugita}, M., {Futamase}, T., {Kasai}, M., \& {Turner}, E.~L. 1992, \apj,
  393, 3

\bibitem[{Hannuksela {et~al.}(2019)Hannuksela, Haris, Ng, Kumar, Mehta, Keitel,
  Li, \& Ajith}]{hannuksela2019search}
Hannuksela, O., Haris, K., Ng, K., {et~al.} 2019, arXiv preprint
  arXiv:1901.02674

\bibitem[{{Haris} {et~al.}(2018){Haris}, {Mehta}, {Kumar}, {Venumadhav}, \&
  {Ajith}}]{Haris2018}
{Haris}, K., {Mehta}, A.~K., {Kumar}, S., {Venumadhav}, T., \& {Ajith}, P.
  2018, arXiv e-prints [\eprint[arXiv]{1807.07062}]

\bibitem[{{Hilbert} {et~al.}(2008){Hilbert}, {White}, {Hartlap}, \&
  {Schneider}}]{Hilbert2008}
{Hilbert}, S., {White}, S.~D.~M., {Hartlap}, J., \& {Schneider}, P. 2008,
  \mnras, 386, 1845

\bibitem[{{Iyer} {et~al.}(2011)}]{M1100296}
{Iyer}, B. {et~al.} 2011, {LIGO India}, Tech. Rep. LIGO-M1100296,
  https://dcc.ligo.org/LIGO-M1100296/public

\bibitem[{{Jung} \& {Shin}(2019)}]{Jung2019}
{Jung}, S. \& {Shin}, C.~S. 2019, Physical Review Letters, 122, 041103

\bibitem[{{Kayser} {et~al.}(1986){Kayser}, {Refsdal}, \&
  {Stabell}}]{Kayser1986}
{Kayser}, R., {Refsdal}, S., \& {Stabell}, R. 1986, \aap, 166, 36

\bibitem[{{Kelly} {et~al.}(2018){Kelly}, {Diego}, {Rodney}, {Kaiser},
  {Broadhurst}, {Zitrin}, {Treu}, {P{\'e}rez-Gonz{\'a}lez}, {Morishita},
  {Jauzac}, {Selsing}, {Oguri}, {Pueyo}, {Ross}, {Filippenko}, {Smith},
  {Hjorth}, {Cenko}, {Wang}, {Howell}, {Richard}, {Frye}, {Jha}, {Foley},
  {Norman}, {Bradac}, {Zheng}, {Brammer}, {Benito}, {Cava}, {Christensen}, {de
  Mink}, {Graur}, {Grillo}, {Kawamata}, {Kneib}, {Matheson}, {McCully},
  {Nonino}, {P{\'e}rez-Fournon}, {Riess}, {Rosati}, {Schmidt}, {Sharon}, \&
  {Weiner}}]{Kelly2018}
{Kelly}, P.~L., {Diego}, J.~M., {Rodney}, S., {et~al.} 2018, Nature Astronomy,
  2, 334

\bibitem[{{Kochanek}(1996)}]{Kochanek1996}
{Kochanek}, C.~S. 1996, \apj, 466, 638

\bibitem[{{Kochanek}(2004)}]{Kochanek2004}
{Kochanek}, C.~S. 2004, \apj, 605, 58

\bibitem[{{Lai} {et~al.}(2018){Lai}, {Hannuksela}, {Herrera-Mart{\'{\i}}n},
  {Diego}, {Broadhurst}, \& {Li}}]{Lai2018}
{Lai}, K.-H., {Hannuksela}, O.~A., {Herrera-Mart{\'{\i}}n}, A., {et~al.} 2018,
  \prd, 98, 083005

\bibitem[{Li {et~al.}(2018)Li, Mao, Zhao, \& Lu}]{li2018gravitational}
Li, S.-S., Mao, S., Zhao, Y., \& Lu, Y. 2018, Monthly Notices of the Royal
  Astronomical Society, 476, 2220

\bibitem[{{Li} {et~al.}(2018){Li}, {Tang}, {Lin}, \& {Wang}}]{Li2018}
{Li}, X., {Tang}, L., {Lin}, H.-N., \& {Wang}, L.-L. 2018, Chinese Physics C,
  42, 095104

\bibitem[{{Liao} {et~al.}(2017){Liao}, {Fan}, {Ding}, {Biesiada}, \&
  {Zhu}}]{Liao2017}
{Liao}, K., {Fan}, X.-L., {Ding}, X., {Biesiada}, M., \& {Zhu}, Z.-H. 2017,
  Nature Communications, 8, 1148

\bibitem[{{Liu} {et~al.}(2018){Liu}, {Guo}, \& {Cai}}]{Liu2018}
{Liu}, L., {Guo}, Z.-K., \& {Cai}, R.-G. 2018, arXiv e-prints
  [\eprint[arXiv]{1812.05376}]

\bibitem[{{Liu} {et~al.}(2019){Liu}, {Guo}, \& {Cai}}]{Liu2019}
{Liu}, L., {Guo}, Z.-K., \& {Cai}, R.-G. 2019, arXiv e-prints
  [\eprint[arXiv]{1901.07672}]

\bibitem[{{Madau} \& {Dickinson}(2014)}]{Madau2014}
{Madau}, P. \& {Dickinson}, M. 2014, \araa, 52, 415

\bibitem[{Maggiore(2008)}]{maggiore}
Maggiore, M. 2008, Gravitational Waves: Volume 1: Theory and Experiments,
  Vol.~1 (Oxford university press)

\bibitem[{{Morishita} {et~al.}(2017){Morishita}, {Abramson}, {Treu}, {Schmidt},
  {Vulcani}, \& {Wang}}]{Morishita2017}
{Morishita}, T., {Abramson}, L.~E., {Treu}, T., {et~al.} 2017, \apj, 846, 139

\bibitem[{Nakamura(1998)}]{nakamura1998gravitational}
Nakamura, T.~T. 1998, Physical review letters, 80, 1138

\bibitem[{{Nakamura}(1998)}]{Nakamura1998}
{Nakamura}, T.~T. 1998, Physical Review Letters, 80, 1138

\bibitem[{{Nakamura} \& {Deguchi}(1999)}]{Nakamura1999}
{Nakamura}, T.~T. \& {Deguchi}, S. 1999, Progress of Theoretical Physics
  Supplement, 133, 137

\bibitem[{{Ng} {et~al.}(2018){Ng}, {Wong}, {Broadhurst}, \& {Li}}]{Ng2018}
{Ng}, K.~K.~Y., {Wong}, K.~W.~K., {Broadhurst}, T., \& {Li}, T.~G.~F. 2018,
  \prd, 97, 023012

\bibitem[{{Oguri}(2018)}]{Oguri2018b}
{Oguri}, M. 2018, \mnras, 480, 3842

\bibitem[{{Oguri} {et~al.}(2018){Oguri}, {Diego}, {Kaiser}, {Kelly}, \&
  {Broadhurst}}]{Oguri2018}
{Oguri}, M., {Diego}, J.~M., {Kaiser}, N., {Kelly}, P.~L., \& {Broadhurst}, T.
  2018, \prd, 97, 023518

\bibitem[{{Paczynski}(1986)}]{Paczynski1986}
{Paczynski}, B. 1986, \apj, 301, 503

\bibitem[{Punturo {et~al.}(2010)Punturo, Abernathy, Acernese, Allen, Andersson,
  Arun, Barone, Barr, Barsuglia, Beker, {et~al.}}]{3gdetectors1}
Punturo, M., Abernathy, M., Acernese, F., {et~al.} 2010, Classical and Quantum
  Gravity, 27, 194002

\bibitem[{{Schneider} {et~al.}(1992){Schneider}, {Ehlers}, \&
  {Falco}}]{SchneiderBook1992}
{Schneider}, P., {Ehlers}, J., \& {Falco}, E.~E. 1992 (Springer-Verlag Berlin
  Heidelberg New York), 112

\bibitem[{{Sereno} {et~al.}(2010){Sereno}, {Sesana}, {Bleuler}, {Jetzer},
  {Volonteri}, \& {Begelman}}]{Sereno2010}
{Sereno}, M., {Sesana}, A., {Bleuler}, A., {et~al.} 2010, Physical Review
  Letters, 105, 251101

\bibitem[{Smith {et~al.}(2018{\natexlab{a}})Smith, Berry, Bianconi, Farr,
  Jauzac, Massey, Richard, Robertson, Sharon, Vecchio,
  {et~al.}}]{smith2018strong}
Smith, G.~P., Berry, C., Bianconi, M., {et~al.} 2018{\natexlab{a}}, arXiv
  preprint arXiv:1803.07851

\bibitem[{Smith {et~al.}(2018{\natexlab{b}})Smith, Bianconi, Jauzac, Richard,
  Robertson, Berry, Massey, Sharon, Farr, \& Veitch}]{smith2018deep}
Smith, G.~P., Bianconi, M., Jauzac, M., {et~al.} 2018{\natexlab{b}}
  [\eprint[arXiv]{1805.07370}]

\bibitem[{{Smith} {et~al.}(2018){Smith}, {Jauzac}, {Veitch}, {Farr}, {Massey},
  \& {Richard}}]{Smith2018}
{Smith}, G.~P., {Jauzac}, M., {Veitch}, J., {et~al.} 2018, \mnras, 475, 3823

\bibitem[{Smith {et~al.}(2018)Smith, Jauzac, Veitch, Farr, Massey, \&
  Richard}]{Smith:2017mqu}
Smith, G.~P., Jauzac, M., Veitch, J., {et~al.} 2018, Mon. Not. Roy. Astron.
  Soc., 475, 3823

\bibitem[{Somiya(2012)}]{somiya:2012detector}
Somiya, K. 2012, Classical and Quantum Gravity, 29, 124007

\bibitem[{Takahashi \& Nakamura(2003)}]{takahashi2003gravitational}
Takahashi, R. \& Nakamura, T. 2003, The Astrophysical Journal, 595, 1039

\bibitem[{{Takahashi} \& {Nakamura}(2003)}]{Takahashi2003}
{Takahashi}, R. \& {Nakamura}, T. 2003, \apj, 595, 1039

\bibitem[{{Takahashi} {et~al.}(2011){Takahashi}, {Oguri}, {Sato}, \&
  {Hamana}}]{Takahashi2011}
{Takahashi}, R., {Oguri}, M., {Sato}, M., \& {Hamana}, T. 2011, \apj, 742, 15

\bibitem[{{The LIGO Scientific Collaboration} {et~al.}(2018){The LIGO
  Scientific Collaboration}, {the Virgo Collaboration}, {Abbott}, {Abbott},
  {Abbott}, {Abraham}, {Acernese}, {Ackley}, {Adams}, {Adhikari}, \&
  et~al.}]{LIGO2018}
{The LIGO Scientific Collaboration}, {the Virgo Collaboration}, {Abbott},
  B.~P., {et~al.} 2018, arXiv e-prints [\eprint[arXiv]{1811.12907}]

\bibitem[{{Turner} {et~al.}(1984){Turner}, {Ostriker}, \& {Gott}}]{Turner1984}
{Turner}, E.~L., {Ostriker}, J.~P., \& {Gott}, III, J.~R. 1984, \apj, 284, 1

\bibitem[{{Ulmer} \& {Goodman}(1995)}]{UlmerGoodman1995}
{Ulmer}, A. \& {Goodman}, J. 1995, \apj, 442, 67

\bibitem[{{Venumadhav} {et~al.}(2017){Venumadhav}, {Dai}, \&
  {Miralda-Escud{\'e}}}]{Venumadhav2017}
{Venumadhav}, T., {Dai}, L., \& {Miralda-Escud{\'e}}, J. 2017, \apj, 850, 49

\bibitem[{{Wambsganss} {et~al.}(1990){Wambsganss}, {Paczynski}, \&
  {Schneider}}]{Wambsganss1990}
{Wambsganss}, J., {Paczynski}, B., \& {Schneider}, P. 1990, \apjl, 358, L33

\bibitem[{{Wang} {et~al.}(1996){Wang}, {Stebbins}, \& {Turner}}]{Wang1996}
{Wang}, Y., {Stebbins}, A., \& {Turner}, E.~L. 1996, Physical Review Letters,
  77, 2875

\bibitem[{{Wyithe} \& {Turner}(2001)}]{Wyithe2001}
{Wyithe}, J.~S.~B. \& {Turner}, E.~L. 2001, \mnras, 320, 21

\bibitem[{{Yang} {et~al.}(2018){Yang}, {Hu}, {Cai}, \& {Wang}}]{Yang2018}
{Yang}, T., {Hu}, B., {Cai}, R.-G., \& {Wang}, B. 2018, arXiv e-prints
  [\eprint[arXiv]{1810.00164}]

\end{thebibliography}


\end{document}